\documentclass[10pt,twocolumn,showpacs,preprintnumbers,amsmath,amssymb,floatfix]{revtex4}

\usepackage{graphicx}
\usepackage{dcolumn} 
\usepackage{bm}      

\def\RotII{{\mathcal R}(\gamma_1,\gamma_2,\gamma_3)}

\def\bn{{\bf n}}

\def\bk{{\bf k}}
\def\br{{\bf r}}
\def\bx{{\bf x}}
\def\bs{{\bf s}}
\def\by{{\bf y}}

\def\bu{{\bf u}}
\def\bq{{\bf q}}

\def\g{{\rm g}}

\def\abc2{{\left(\frac{ab}{c^2}\right)}}

\def\WS{{\rm WS}}

\def\nbarh{\bar{n}_{\rm H}}

\def\nbar{\bar{n}}

\def\dx{{d^3\!x}}

\def\dk{{d^3\!k}}
\def\dy{{d^3\!y}}
\def\dq{{d^3\!q}}

\def\tpi3{(2\pi)^3}
\def\fom{f(\Omega)}

\def\be{\begin{equation}}
\def\ee{\end{equation}}
\def\rhob{\bar{\rho}}

\def\nbar{\bar{n}}
\def\simless{\mathbin{\lower 3pt\hbox
  {$\rlap{\raise 5pt\hbox{$\char'074$}}\mathchar"7218$}}}   
\def\simgreat{\mathbin{\lower 3pt\hbox
   {$\rlap{\raise 5pt\hbox{$\char'076$}}\mathchar"7218$}}}  

\def\ba{\begin{eqnarray}}
\def\ea{\end{eqnarray}}

\def\ST{\rm ST}
\def\vir{\rm vir}
\def\bess0{{\rm J_0}}
\def\sbess0{{\rm j_0}}
\def\cyc{{\rm cyc}}
\def\1H{1{\rm H}}
\def\2H{2{\rm H}}
\def\3H{3{\rm H}}

\def\kperp{{\bf k_{\perp}}}
\def\Mpc{{\rm Mpc}}

\def\CgI{\cos{\gamma_1}\,}
\def\SgI{\sin{\gamma_1}\,}
\def\CgII{\cos{\gamma_2}\,}
\def\SgII{\sin{\gamma_2}\,}

\def\Vu{V_{\mu}}

\def\psidij{\psi_{\delta,j}^{(i)}(\bk_1,\dots,\bk_j)}
\def\psivj{\psi_{v,j}(\bk_1,\dots,\bk_j)}

\def\psid{\psi_{\delta,j}^{(i)}}
\def\psiv{\psi_{v,j}}

\def\hc{\rm hc}

\def\erf{\rm erf}

\def\D1{\rm 1D}

 \def\ga{\mathrel{\mathpalette\fun >}}
 \def\fun#1#2{\lower3.6pt\vbox{\baselineskip0pt\lineskip.9pt
        \ialign{$\mathsurround=0pt#1\hfill##\hfil$\crcr#2\crcr\sim\crcr}}}

\begin{document}

\title{An analytic model for the bispectrum of galaxies in redshift space}

\author{Robert E. Smith$^{1,2}$, Ravi K. Sheth$^{1}$ and Rom\'an
Scoccimarro$^{3}$ \small \\ \vspace{0.3cm}
(1) University of Pennsylvania, 209 South 33rd Street, Philadelphia,
    PA 19104, USA. \\
(2) Institute for Theoretical Physics, University of Zurich,
    Zurich, CH8051, Switzerland \\
(3) CCPP, Department of Physics, New York University, New York, NY 10003, USA.  \\
email: {\tt res@physik.unizh.ch, shethrk@physics.upenn.edu,
rs123@nyu.edu, }}

\date{\today}

\vspace{0.1cm}


\begin{abstract}
We develop an analytic theory for the redshift space bispectrum of
dark matter, haloes and galaxies. This is done within the context of
the halo model of structure formation, as this allows for the
self-consistent inclusion of linear and non-linear redshift space
distortions and also for the non-linearity of the halo bias.  The
model is applicable over a wide range of scales: on the largest scales
the predictions reduce to those of the standard perturbation theory
(PT); on smaller scales they are determined primarily by the nonlinear
virial velocities of galaxies within haloes, and this gives rise to
the \emph{U}-shaped anisotropy in the reduced bispectrum -- a finger
print of the Finger-Of-God distortions.  We then confront the
predictions with measurements of the redshift space bispectrum of dark
matter from an ensemble of numerical simulations. On very large
scales, $k=0.05\,h\Mpc^{-1}$, we find reasonably good agreement
between our Halo Model, PT and the data, to within the errors. On
smaller scales, $k=0.1\,h\Mpc^{-1}$, the measured bispectra differ
from the PT at the level of $\sim10-20\%$, especially for colinear
triangle configurations. The Halo Model predictions improve over PT,
but are accurate to no better than $10\%$. On smaller scales
$k=0.5-1.0\,h\Mpc^{-1}$, our model provides a significant improvement
over PT, which breaks down.  This implies that studies which use the
lowest order PT to extract galaxy bias information are not robust on
scales $k\ga 0.1\,h\Mpc^{-1}$.  The analytic and simulation results
also indicate that there is \emph{no} observable scale for which the
configuration dependence of the reduced bispectrum is
constant---hierarchical models for the higher order correlation
functions in redshift space are unlikely to be useful. It is hoped
that our model will facilitate extraction of information from
large-scale structure surveys of the Universe, because different
galaxy populations are naturally included into our description.
\end{abstract}

\pacs{98.80.-k}

\maketitle


\section{Introduction}

Statistical analyses of the large scale structures observed in galaxy
surveys can provide a wealth of information about the cosmological
parameters, the underlying mass distribution and the initial
conditions of the
Universe\cite{Tegmarketal2004,Coleetal2005,Eisensteinetal2005,Tegmarketal2006,Spergeletal2007}. Some
complex combination of the information is commonly extracted through
measurement of the 2-point correlation function, or it's Fourier space
analogue the power spectrum. Since the evolved density field of
galaxies is highly non-Gaussian, further complementary information is
contained within the higher order clustering statistics
\cite{Bernardeauetal2002,SeSc2005}. For example, analysis of the large-scale
3-point correlation function, or its Fourier space dual the
bispectrum, on large scales, can: test the non-linearity of bias --
the way in which an observable tracer distribution samples the
unobservable distribution of physical interest
\cite{FryGaztanaga1993,FriGaz1994,Fry1994,Sefusattietal2006};
constrain our hypothesis of Gaussianity in the initial conditions
\cite{FryScherrer1994,GazFos1998,ScSeZa2004,SefusattiKomatsu2007};
break degeneracies between parameters, hence allowing improved
constraints on the amplitude of the matter power spectrum. In addition,
higher order statistics have been highlighted as an important
piece of solving the puzzle as to whether the observed accelerated
expansion of the Universe is due to Dark Energy physics or a
modification to gravity
\cite{Bernardeau2004,SSHSY2007,JainZhang2007,Scoccimarro2008}. On
smaller scales these statistics can be most usefully used as a
discriminator for the shapes of haloes \cite{SmithWattsSheth2006} --
and thus have the potential to constrain the small-scale dark matter
physics.

The current state of the art galaxy redshift surveys
\cite{Saundersetal2000,Collessetal2001,Yorketal2000,Schneideretal2007}
have provided large samples of the Universe, and investigators have
already carried out some of the tests noted above: 3-point correlation
functions have been estimated
by~\cite{Kayoetal2004,JingBoerner2004,Wangetal2004,
Gaztanagaetal2005,Nicholetal2006,Kulkarnietal2007} and bispectra by
\cite{IRASbisp,Feldmanetal2001,Verdeetal2002,Nishimichietal2006}.

The recovery of precise cosmological information from these
measurements is not straightforward, owing to the influence of non-linear mass
evolution, biasing and redshift space distortions.  Current theoretical modeling of the large
scale bispectrum rests on results from perturbation theory and
non-linear biasing in real (as opposed to redshift) space. In the
large scale limit, this gives rise to a simple model
\cite{FryGaztanaga1993,FriGaz1994,Fry1994}:
\be Q^{\rm g}(k_1,k_2,\theta_{12}) =\frac{1}{b^{\g}_1}\left[Q^{\rm
m}(k_1,k_2,\theta_{12})+c^{\g}_2 \right]\ ;\label{eq:PTs}\ee
where the functions $Q^i$ are the reduced bispectra of galaxies and
matter, $Q^i_{123}\equiv
B^i_{123}/\left[P^i_1P^i_2+P^i_2P^i_3+P^i_3P^i_1\right]$. In the above
$B_{123}\equiv B(k_1,k_2,\theta_{12})$ is the matter bispectrum and
$P_a\equiv P(k_a)$ the matter power spectrum. The coefficient
$b^{\g}_1$ is the large-scale linear bias parameter and
$c^{\g}_2\equiv b_2^{\g}/b_1^{\g}$ is the first non-linear bias
parameter.  It is usually assumed that this relation holds in redshift
space as well, but that it does not in detail can be seen from the
work of \cite{Scoccimarroetal1999}.  That it nevertheless appears to
be a reasonable working hypothesis was demonstrated by
\cite{GaztanagaScoccimarro2005}.  We shall, hereafter, refer to this
model in real and redshift space as PT and PTs, respectively.

There are several reasons why we wish to improve upon PTs.  Firstly,
it is well known that in redshift space the distortion effects from
non-linear structures, such as Finger-of-God (hereafter FOG)
distortions, pollute the scales that are usually identified for linear
treatment.  This can not be accounted for in the perturbation theory
in any other way than supposing `ad-hoc' fixes to the model
\cite{Verdeetal1998,Scoccimarroetal1999}.  A pragmatist might argue
that one may take scales that are sufficiently large that these
corrections can be neglected. However, even if we are proficient
enough to accurately separate linear from non-linear scales, then we
are still faced with loosing a significant amount of information from
our data through the restrictions to very large scales. Therefore some
means for robustly modeling the FOG effects is clearly of great value
as this may allow us to expand the utility of our data set and improve
precision.

Secondly, in our study of the large-scale galaxy power spectrum
\cite{SmithScoccimarroSheth2007} we found that there was non-trivial
scale dependence arising from non-linear bias and gravitational mode
coupling, even on the largest scales currently probed. One may then
ask how these properties affect the predicted bispectra.

Thirdly, if we assume that that galaxy velocity field is an unbiased
tracer for the velocity field of dark matter, then through studying
the higher order clustering statistics in redshift space we have a
direct probe of the statistical information of the dynamics of the CDM
density field itself
\cite{DavisPeebles1983,Kaiser1987,Hamilton1998,Scoccimarro2004}.

In this paper we build a new analytic model for the fully non-linear
redshift space bispectrum. We will concentrate on the isotropically averaged 
(i.e. the monopole) bispectrum, for reasons of simplicity. 
We work in the context of the halo model
\cite{CooraySheth2002}, since it naturally affords a means for
including linear and non-linear density and velocity information
\cite{Shethetal2001,White2001,Seljak2001,Kangetal2002,Smithetal2008}
and neatly allows for the inclusion of galaxies
\cite{Bensonetal2000,Seljak2000,PeacockSmith2000,Scoccimarroetal2001,Shethetal2001,BerlindWeinberg2002}.
Furthermore, as was shown in \cite{SmithScoccimarroSheth2007} the
halo model presents a natural framework for understanding the origins
of the non-linear scale dependence of bias. However, the limitations
of the halo model predictions for precision measurements of the matter
power spectrum on large scales have been known for some time now
\cite{CooraySheth2002,Smithetal2003,CrocceScoccimarro2008,NeyrinckSzapudi2008}.  We will
therefore use measurements from numerical simulations, to confirm the
validity of our predictions.

The paper breaks up as follows: In Section \ref{sec:HaloModel} we
formalize the halo model in redshift space, providing general
expressions for the 3-point function and bispectrum. Section
\ref{sec:functions} details the necessary components of the model; we
pay special attention to the redshift-space clustering of halo
centers.  Section \ref{sec:IsoBispec} presents the central analytic
result of the paper---a calculation of the  bispectrum monopole.  Some
results of evaluating our expressions are presented in Section
\ref{sec:results}. In Section \ref{sec:simulations} we confront our
model with measurements from $N$-body simulations. In Section
\ref{sec:conclusions} we summarize our conclusions.

Although our analysis is general, we shall illustrate our results with
specific examples.  When necessary, we assume a flat
Friedmann-Lema\^itre-Robertson-Walker (FLRW) cosmological model with
energy density at late times dominated by a cosmological constant
($\Lambda$) and a sea of collisionless cold dark matter particles as
the dominant mass density.  We set $\Omega_m=0.27$ and
$\Omega_{\Lambda}=0.73$, where these are the ratios of the energy
density in matter and a cosmological constant to the critical density,
respectively. We use a linear theory power spectrum generated from
{\ttfamily cmbfast}\cite{SeljakZaldarriaga1996}, with baryon content
of $\Omega_b =0.046$ and $h=0.72$.  The normalization of fluctuations
is set through $\sigma_8=0.9$, which is the r.m.s. variance of
fluctuations in spheres of radius $8 h^{-1}\,\Mpc$.


\section{Halo model in redshift space}\label{sec:HaloModel}

\subsection{Formalism}\label{ssec:formalism}

In the halo model (see \cite{CooraySheth2002} for a review) the
density field is decomposed into a set of dark matter haloes, where a
halo is defined to be a region that has undergone gravitational
collapse forming a dense virialized ball of cold dark matter
(CDM). All statistical quantities of interest are then considered as
sums over the halo distribution.  Thus to understand the large scale
clustering of a distribution of objects, haloes, galaxies or dark
matter, we simply require understanding of how the haloes themselves
cluster; the different tracer types simply act as weights.  In
particular, different galaxy populations `weight' haloes differently:
the Halo Occupation Distribution (HOD)
\cite{Bensonetal2000,Seljak2000,PeacockSmith2000,Scoccimarroetal2001,
BerlindWeinberg2002}, specifies how the probability for obtaining $N$
galaxies depends on halo mass $M$.  To model redshift space
statistics, we require additional information about how the large
scale velocity field modifies the halo clustering, as well as a model
for the distribution function of galaxy velocities within each halo.
The halo model in redshift space, at the 2-point level, was developed
by \cite{White2001,Seljak2001,Kangetal2002} (hereafter we shall refer
to the Halo model in real and redshift space as HM and HMs,
respectively). However, some unresolved issues remained with regard to
the base formalism. These were resolved by \cite{Smithetal2008} and
our description of 3-point statistics presented here extends these
analyses. For completeness some of these details are repeated
below. Before continuing, we note that the problem of redshift space
distortions in the Halo Model was also recently addressed by
\cite{Tinker2007}, who used numerical simulations to construct an
empirical model for the distribution function of halo pair-wise
velocities. Our approach is complimentary to that, since the results
for the large-scale halo clustering are derived within the context of
the analytic perturbation theory rather than being fit for.

The density field of dark matter, haloes or galaxies may be written as
\be \rho^s_{\alpha}(\bs) = \sum_i \left[W_{\alpha}\right]_i
                            U^s_{\alpha,i}(\bs-\bs_i|M_i)\ ,
\label{eq:HaloModRed}\ee
where $\alpha=\{1,2,3\}$ refers to the particular choice of weight for
the $i$th halo in the sum,
i.e. $[W_{\alpha}]_i=\{1,M_i,N^{\g}(M_i),\dots\}$ depending on the
spectra one wishes to model, and where $N^{\g}(M_i)$ is the number of
galaxies in halo $i$. $U^s_{\alpha,i}$ is the normalized density
distribution of objects in redshift space within the $i$th halo. In
this paper we shall always assume that
$U^s_{\alpha,i}\equiv\rho^s(\bs)/M$, is the mass-normalized density
profile of dark matter in redshift space, although our formalism does
not rely upon this assumption and may readily be generalized for more
complicated mass distributions. At this point the only difference
between Eq.~(\ref{eq:HaloModRed}) and the real space density field, is
that we have used $\bs$ to denote comoving spatial positions.  However
this has the special meaning that Hubble's law, ${\bf v} = H(a) {\bf
r}$, is used to infer proper radial positions from recession
velocities, where ${\bf v}$ is the proper velocity, $H(a)\equiv
\dot{a}/a$ is the Hubble parameter and $\br$ is the proper separation
(related to comoving coordinate through $\br=a\,\bx$). The notion of
redshift space distortions then follow from the fact that objects
which form through gravitational instability acquire a local peculiar
velocity of their own, and hence the velocity--space mapping in
general is non-linear.  In this paper we shall work in {\em the plane
parallel approximation}, where observed structures are located at
infinity. Then the mapping is
\be s_{z}=z-u_{z}(\bx)\ ;\ \bs_{\perp}=\bx_{\perp} \, ,\label{eq:RedMap}\ee
where the Cartesian components of the position vectors have been
written $(\bx_{\perp},z)$, with $\bx_{\perp}=(x,y)$. Thus $s_z$ and
$u_{z}$ specify the z-components of the redshift space position vector
$\bs$ and the comoving peculiar velocity field $\bu$, scaled in units
of the Hubble parameter, respectively. Note that we take $\bu$ to be
negative for convenience.


\subsection{Higher order correlations}

We may now compute the correlation hierarchy for such a distribution
of tracer objects.  For a definition of the higher-order clustering
statistics in configuration space and their Fourier space dual
counterparts we refer to Appendix \ref{sec:theory}. There may also be
found useful symmetry properties that we exploit throughout.

Following
\cite{ScherrerBertschinger1991,Scoccimarroetal2001,CooraySheth2002,TakadaJain2003,SmithWattsSheth2006},
the real-space 3-point correlation function ($\zeta_{\alpha}^{s}$) in
the halo model, for dark matter, haloes or galaxies, is the sum of
three terms: the first represents the case where all three points in
space are contained in a single halo; the second is the case where two
points are located in one halo and the third is in a separate halo;
the third is the case where three points are located in three distinct
haloes -- we shall refer to these as the 1-, 2- and 3-Halo terms and
($\zeta^{s}_{\alpha,\rm 1H}$, $\zeta^{s}_{\alpha,\rm 2H}$,
$\zeta^s_{\alpha,\rm 3H}$). These are written:
\begin{widetext}
\ba \zeta^s_{\alpha}(\bs_{1},\bs_{2},\bs_3) & \equiv &
\zeta^s_{\alpha,\1H}(\bs_{1},\bs_{2},\bs_3) +
\zeta^s_{\alpha,\2H}(\bs_{1},\bs_{2},\bs_3) +
\zeta^s_{\alpha,\3H}(\bs_{1},\bs_{2},\bs_3) \ \label{eq:RedXi3}\ ;\\
\zeta^s_{\alpha,\1H}(\bs_{1},\bs_{2},\bs_3) & = & 
 \frac{1}{\rhob^3_{\alpha}} \int dM\; \dy \;[W_{\alpha}]^3 \; n(M)
 \prod_{i=1}^{3}\left\{\frac{}{} U^s(\by-\bs_{i}|M)\right\}\;\label{xi3-1H}\ ;\\
\zeta^s_{\alpha,\2H}(\bs_{1},\bs_{2},\bs_3) & = & 
 \frac{1}{\rhob_{\alpha}^3}\int \!\!\prod_{i=\{1,2\}}\!\!  
 \left\{ \frac{}{}dM_i\, \dy_i\, [W_{\alpha}]_i\, n(M_i) \ U^s(\by_i-\bs_i|M_i) \right\} 
 \,[W_{\alpha}]_1 \ U^s(\by_1-\bs_3|M_1)\nonumber \\
& & \qquad\qquad\qquad \times \ \;
\xi^s_{\hc}(\by_1,\by_2|M_1,M_2)+\cyc\ ;\label{xi3-2H} \\
\zeta^s_{\alpha,\3H}(\bs_{1},\bs_{2},\bs_3) & = & 
\frac{1}{\rhob_{\alpha}^3}\int \prod_{i=1}^{3}
\left\{ \frac{}{} dM_i\; \dy_i\; [W_{\alpha}]_i\, n(M_i)\, U^s(\by_i-\bs_i|M_i)\right\} 
\zeta_{\hc}^s(\by_1,\by_2,\by_3|M_1,M_2,M_3)\ \label{xi3-3H} \ ,\ea
\end{widetext}
where $\xi_{\hc}^s$ and $\zeta_{\hc}^s$ are the 2- and 3-point
correlation functions of halo centers, conditioned on halo masses and
where $n(M)dM$ is the halo mass function, which gives the number
density of dark matter haloes with masses in the range $M$ to $M+dM$.

The inverse Fourier transforms of these 3-point functions are the
redshift space bispectra (c.f. Eq.~\ref{eq:Bi2Cf3}).  They are
written:
\begin{widetext}
\ba 
B^s_{\alpha}(\bk_1,\bk_2,\bk_3) & = & 
B^s_{\alpha,\1H}(\bk_1,\bk_2,\bk_3)+
B^s_{\alpha,\2H}(\bk_1,\bk_2,\bk_3)+
B^s_{\alpha,\3H}(\bk_1,\bk_2,\bk_3)\ ,\\
B^s_{\alpha,\1H}(\bk_1,\bk_2,\bk_3) & = & \frac{1}{\rhob^3_{\alpha}}
 \int dM\, {[W_{\alpha}]}^3\,n(M) \prod_{i=1}^{3} \left\{\frac{}{}U^s(\bk_i|M)\right\}
 \ , \label{eq:bi1H}\\
B^s_{\alpha,\2H}(\bk_1,\bk_2,\bk_3) & = & \frac{1}{\rhob^3_{\alpha}} \int \prod_{i=\{1,2\}}
 \left\{\frac{}{}dM_i\, [W_{\alpha}]_i\,n(M_i) U^s(\bk_i|M_i) \right\}
 [W_{\alpha}]_1\, U^s(\bk_3|M_1) P^s_{\hc}(\bk_2|M_1,M_2) + \cyc\
,\label{eq:bi2H} \\
B^s_{\alpha,\3H}(\bk_1,\bk_2,\bk_3) & = & \frac{1}{\rhob^3_{\alpha}}\int \prod_{i=1}^{3}
\left\{\frac{}{}dM_i \,[W_{\alpha}]_i\, n(M_i)\,U^s(\bk_i|M_i)\right\}
B^s_{\hc}(\bk_1,\bk_2,\bk_3|M_1,M_2,M_3) \
,\label{eq:bi3H}\ \ea
\end{widetext}
where $P^{s}_{\hc}$ and $B^s_{\hc}$ are the Fourier transforms of the
2- and 3-point halo center correlation functions.

Several advantages are gained from transforming to Fourier space.
Firstly, once the integrals over mass are included, $\zeta^s_{\rm 3H}$
requires evaluation of a 12-D integral---the corresponding term
$B^s_{\3H}$ is significantly simpler.  Indeed, for the case of real
space -- not redshift space -- and for spherical haloes, it is
possible to write $B_{\3H}$ as the product of 3 2-D integrals. The
calculation is slightly more complicated in redshift space but, as we
show below, it remains tractable.  Thus, to compute the redshift space
power spectrum and bispectrum, we require three components: the
abundance of dark matter haloes $n(M)$; a model for the redshift space
density profile; and a model for the inter-clustering of dark matter
haloes in redshift space. In the following sections we describe our
choices for these quantities.


\section{Ingredients}\label{sec:functions}

\subsection{Halo abundances and bias factors}\label{sec:nM}

The halo mass function $n(M)$ plays a central role in the halo model.
It has been the subject of much detailed study
\cite{PressSchechter1974,Bondetal1991,ShethTormen1999,ShethMoTormen2001}.
These studies suggest that, in appropriately scaled units, halo
abundances should be approximately independent of cosmology, power
spectrum and redshift.  These models for $n(M)$ also predict that the
real-space clustering of halos should be biased relative to that of
the dark matter \cite{ShethTormen1999}; the way in which real-space
halo bias depends on halo mass is related to the shape of $n(M)$.
Thus, once the mass function has been specified, the problem of
describing halo clustering reduces to one of describing the clustering
of the dark matter.  Of the many recent parametrizations of $n(M)$,
\cite{ShethTormen1999,Jenkinsetal2001,Warrenetal2006,Reedetal2007}, we
use that of Sheth \& Tormen \cite{ShethTormen1999}. Changing to that
of Warren et al. \cite{Warrenetal2006} for instance, does not affect
the large-scale matter predictions, and changes the results in the
non-linear regime by a few percent. Note that for the bispectrum, as
was shown by \cite{Scoccimarroetal2001}, a more important issue to be
aware of is the finite volume effect, which can significantly change
the measured statistics for small volumes.  The halo bias factors
associated with this mass function are reported in
\cite{Scoccimarroetal2001}; we use these in what follows.
\cite{SmithScoccimarroSheth2007} describe other empirical approaches
to determining halo bias parameters.


\subsection{Density profiles in redshift space}\label{sec:RedDen}

Consider the 6-D phase space density distribution function for dark
matter particles within a particular halo, denoted
$\mathcal{F}(\bx,\bu|M)$.  The density profile and velocity
distribution function may be obtained by marginalizing over velocities
and positions, respectively:
\be \rho(\bx)=M\int d\bu\, \mathcal{F}(\bx,\bu|M)\ ; \hspace{0.2cm}
{\mathcal V}_{\rm 3D}(\bu)=\int d\bx\, \mathcal{F}(\bx,\bu|M) , \ee
where $M$ is the normalizing mass.  The redshift space density profile
can be readily obtained from the phase space distribution through
transformation to the new random variable $\bs$, given by our
fundamental mapping (Eq.~\ref{eq:RedMap}). Hence
\ba \rho^s(\bs_{\perp},s_z) & = & M\,\int dz\,d\bx_{\perp} \, du_z\,
 d\bu_{\perp}\, \mathcal{F}(\bx_{\perp},z,\bu_{\perp},u_z|M) \nonumber \\
& & \hspace{0.4cm}\times \delta^D(s_z-z+u_z) \,\delta^D(\bs_{\perp}-\bx_{\perp})\nonumber ,\\
 & = & M\, \int dz\, d\bu_{\perp}\, 
       \mathcal{F}(\bs_{\perp},z,\bu_{\perp},z-s_z|M) \ .
\ea
We now assume that the density distribution of matter within each halo
is well described by a spherically symmetric density profile and that
the particle orbits are isotropic and independent of position within
the halo.  Thus, the phase space distribution is separable, i.e.
$\mathcal{F}(\bx,\bu|M)=\rho(\bx|M){\mathcal V}_{\rm 3D}(\bu|M)/M$.
Since the velocity distribution function is isotropic it may now be
written as the product of three independent distributions in the three
coordinate directions: ${\mathcal V}_{\rm 3D}(\bu|M)={\mathcal V}_{\rm
1D}(u_x|M){\mathcal V}_{\rm 1D}(u_y|M){\mathcal V}_{\rm 1D}(u_z|M)$;
and we shall hereafter use the notation that ${\mathcal V}_{\rm
1D}\equiv{\mathcal V}$.  Hence,
\be \rho^s(\bs_{\perp},s_z|M) = \int dz\, \rho(\bs_{\perp},z|M)\,
  {\mathcal V}(s_z-z|M) .\ee
Fourier transforming yields the compact expression
\be U^s(\kperp,k_z|M) = U(k|M)\, {\mathcal V}(\mu k|M) \ ,
\label{eq:redprofile}\ee
where $\kperp$ denotes a 2-D wavevector perpendicular to the
distortion, and $k_z=\mu k=\bk\cdot\hat{\bf z}$ is parallel to it.

This expression shows that the redshift space profile is anisotropic
because the spherically symmetric real-space profile $U$ has been
convolved along the line-of-sight direction with displacements
generated by the velocity distribution ${\mathcal V}$.  That is to
say, ${\mathcal V}$ is the quantity in the model which generates FOG
distortions and since it represents virial motions, it is clearly the
sort of nonlinear effect that PTs based approaches must model `ad
hoc'. We also note that if ${\mathcal V}$ makes the redshift profile
substantially anisotropic, then these nonlinear effects may extend
farther into the linear regime than one might otherwise have expected.

We caution that this simple model of the halo phase space will almost
certainly not be strictly valid, since it clearly neglects many
aspects of the more complex physics that we understand to play an
important role for the internal dynamics of haloes.
\footnote{For example: substructures will produce localized features
in the phase space distribution \cite{Mooreetal1999,Klypinetal1999}; a
global asymmetry of the underlying potential, generated through the
anisotropic accretion of matter, will distort the velocity structure
of the phase space into a 6-D triaxial ellipsoid
\cite{JingSuto2002}, etc.}
Nevertheless, results from numerical simulations show that the
isotropic model is a good approximation, which allows one to write down
simple expressions that provide physical insight. 
\footnote{Kang et al \cite{Kangetal2002} show that if one considers
the ensemble average of the phase space distribution for haloes, then
${\mathcal V}(\mu k|M)$ is reasonably well described by a
Maxwellian. This result was further corroborated by
\cite{Kuwabaraetal2002}, who computed ${\mathcal V}(\mu k|M)$ for a
halo with an NFW density profile \cite{NavarroFrenkWhite1997} by
solving the Jeans Equation. They found that it was well approximated
by a Maxwellian.}

Precise details of the models we employ for the density profile of
dark matter and for the 1-Point distribution function of velocities
are presented in Appendix~\ref{sec:details}. Note also that owing to
these models employing different conventions for the halo mass we must
convert between them, and we do this using our procedure from
\cite{SmithWatts2005}, see Appendix \ref{ssec:MassConversion} for
some details.


\subsection{Halo center clustering in redshift space}\label{ssec:RedHaloSeedBispec}

On large scales the success of our analytic model will primarily be
determined by its ability to reproduce the large-scale clustering of
the halo centers.  For this we use the redshift space Halo-PT
developed in \cite{Smithetal2008}, which is accurate up to the 1-loop
level in perturbation theory. The main result we draw from that work
is the idea that the halo density field may be written as perturbation
series that involves the standard density PT kernels
\cite{Bernardeauetal2002} and the non-linear bias parameters
\cite{FryGaztanaga1993}. Explicitly we have the series,
\ba 
\delta^s_{\hc}(\bk,a|M,R) & = & \sum_{n=0}^{\infty}[D_1(a)]^n 
[\delta_{\hc}^s(\bk|M,R)]_n \label{eq:reddenPT1} \\
\left[\delta_{\hc}^s(\bk|M,R)\right]_n & = & \int
\prod_{i=1}^{n}\left\{\frac{\dq_i}{(2\pi)^3}\,\delta_1(\bq_i)\right\}
(2\pi)^3\left[\delta^D(\bk)\right]_n \nonumber \\ 
& & \hspace{0.6cm} \times \ Z^{\hc}_n(\bq_1,...,\bq_n|M,R)
\label{eq:reddenPT2}  ,\ea
where $D(t)$ is the linear theory growth function and $f(\Omega)\equiv
d \log D(a)/d \log a$ is the logarithmic growth rate of the velocity
field. The functions $Z^{\hc}_n(\bq_1,...,\bq_n|M,R)$ are the redshift
space Halo-PT kernels symmetrized in all of their arguments and we
make explicit their dependencies on halo mass and the scale over which
the density field has been smoothed.  Kernels up to second order are
\cite{Hivonetal1995,Verdeetal1998,Scoccimarroetal1999}:
\ba
Z_0^{\hc} & = &  F_0^{\hc}\equiv b_0 \ ;\label{eq:ZPT0}\\
Z_1^{\hc} & = &  F_1^{\hc}+f(\Omega)\mu_{1}^2\, \tilde{G}_1
\left[ 1 + b_0  \right]  \ ;\label{eq:ZPT1}\\
Z^{\hc}_{1,2} & = & 
F^{\hc}_{1,2}+f(\Omega)\mu_{12}^2 \tilde{G}_{1,2}
\nonumber\\
& + & \frac{1}{2}f(\Omega)\mu_{12} k_{12} 
\left\{\frac{\mu_{1}}{q_1}\tilde{G}_1
\left[F_2^{\hc}+f(\Omega)\mu_{2}^2\tilde{G}_2\right] \right. \nonumber \\
& & \left. \hspace{0.4cm} +\frac{\mu_{2}}{q_2}\tilde{G}_2\left[F_1^{\hc}
+f(\Omega)\mu_{1}^2\tilde{G}_1 \right]
\right\} \nonumber \\ 
& + & \frac{1}{2}\left[f(\Omega)\mu_{12} k_{12}\right]^2\frac{\mu_{1}}{q_1}
\frac{\mu_{2}}{q_2}\tilde{G}_1\tilde{G}_2\, b_0 \label{eq:ZPT2}
\ , \ea
where we have adopted the short-hand notation:
\ba
Z^{\hc}_{i_1,\dots,i_n} & \equiv & Z^{\hc}_n(\bq_{i_1},\dots,\bq_{i_n}|M,R)\ ;
\nonumber \\
F^{\hc}_{i_1,\dots,i_n} & \equiv & F^{\hc}_n(\bq_{i_1},\dots,\bq_{i_n}|M,R)\ ;
\nonumber \\
\tilde{G}_{i_1,\dots,i_n} & \equiv & W(\left|\bq_{i_1}+\dots+\bq_{i_n}\right|R)
G_n(\bq_{i_1},\dots,\bq_{i_n})\ ;\nonumber  \ea
and where
\ba \mu_{i_1\dots i_n} & \equiv &
\frac{(\bq_{i_1}+\dots+\bq_{i_n})\cdot \hat{\bf z}}{k_{i_1\dots i_n}} \ ;\\ 
k_{i_1\dots i_n} & \equiv & \left|\bq_{i_1}+\dots+\bq_{i_n}\right|\ .
\label{eq:dircosines}\ea
The quantities $F^{\hc}_n$ are the $n$th order Halo-PT kernels (see
Appendix~\ref{sec:HaloPTKernels} and \cite{SmithScoccimarroSheth2007}
for complete details). The functions $G_n$ represent the $n$th order
Eulerian PT kernels for the divergence of the velocity field
\cite{Bernardeauetal2002}.  Note that these expressions are almost
identical to the redshift space PT kernels derived by
\cite{Scoccimarroetal1999,Verdeetal1998},
however they differ in some subtle ways: one, we have explicitly
included their dependence on the smoothing filter $W(q)$, which is
needed to facilitate the Taylor expansion; and two, we are applying
this in the context of haloes and not galaxies and so they depend on
the non-linear halo bias parameters $b_i(M)$, instead of the nonlinear
galaxy bias parameters (see discussion in Section~\ref{sec:nM}).  Note
that we have also included $b_0(M)$, since this does not have to be
zero, although we will take it to be so for all our later analysis.

Following standard methods for calculating polyspectra, we find that
the halo center bispectrum, $B^s_{\hc,123}\equiv
B^s_{\hc}(\bk_1,\bk_2,\bk_3|M_1,M_2,M_3,R)$, up to fourth order in
redshift space Halo-PT, is
\ba  B^s_{\hc,123} & = &
  2P_{11}(k_1)P_{11}(k_2) \, Z_1^{\hc}(\bk_1|M_1,R)\,
  Z_1^{\hc}(\bk_2|M_2,R) \nonumber \\ 
  & & \times\ Z_{2}^{\hc}(\bk_1,\bk_2|M_3,R) +{2\ \rm cyc
  } \ .\label{eq:RedBiHaloCent}\ea
Inserting our expressions for the redshift space Halo-PT kernels,
Eqs~(\ref{eq:ZPT0}--\ref{eq:ZPT2}), in to the above expression,
reveals:
\begin{widetext}
\ba 
\frac{B^s_{\hc,123}(M_1,M_2,M_3)}{W(k_1R)W(k_2R)W(k_3R)} 
& = & 2b_{1}(M_1)b_1(M_2)b_1(M_3)P_{11}(k_1)P_{11}(k_2) \, \prod_{i=1,2}\left\{
1+\beta_i\,\mu_i^2\left[1+b_0(M_i)\right] \right\}
\nonumber \\
 & & \hspace{-3cm} \times \ 
\left[ 
   F_{2}(\bk_1,\bk_2)+\beta_3\,\mu_{12}^2 G_{2}(\bk_1,\bk_2)
   +\frac{W(k_1R)W(k_2R)}{W(k_3R)}
   \left(
       \frac{c_2(M_3)}{2}+\frac{1}{2}\left[f(\Omega)\mu_{12} {k_{12}}\right]^2
       \frac{\mu_{1}}{k_1}\frac{\mu_{2}}{k_2}c_0(M_3) 
   \right. 
\right.
\nonumber\\
& & \hspace{0cm}
\left. 
   \left.
       + \frac{1}{2}f(\Omega)\mu_{12} k_{12} 
         \left\{\frac{\mu_{1}}{k_1}\left[1+\beta_3\,\mu_{2}^2\right] 
+               \frac{\mu_{2}}{k_2}\left[1+\beta_3\,\mu_{1}^2\right]
         \right\}
   \right)
\right]\ + 2\,\cyc.
\label{eq:RedBiHaloCent2}\ea
\end{widetext}
where $\beta_i\equiv f(\Omega)/b_1(M_i)$ and where
$c_j(M_i)=b_j(M_i)/b_1(M_i)$. As in our real space work on the power
spectrum, we are now faced with the situation that we have solved for
the bispectrum of halo centers filtered on scale $R$, and in fact we
would like to recover the unfiltered bispectrum. As in
\cite{SmithScoccimarroSheth2007}, we take the following {\em ansatz}:
the filtering of the spectra can be reversed through the following
operation:
\ba 
P(\bk) & \equiv & \frac{P(\bk|R)}{W^2(kR)}\ ;\ \nonumber \\
B(\bk_1,\bk_2,\bk_3) & \equiv & \frac{B(\bk_1,\bk_2,\bk_3|R)}{W(k_1R)W(k_2R)W(k_3R)} 
\ea
and this explains the form of the left-hand-side of
Eq.~(\ref{eq:RedBiHaloCent2}).  An alternative approach to the
filtering issue for the power spectrum and bispectrum was proposed by
\cite{McDonald2006}, we shall leave the solution of this problem for future consideration.

\vspace{0.2cm}

With these ingredients prepared, we are now in full possession of a
complete description of the bispectrum of galaxies, haloes and dark
matter in the halo model and in the presence of a local, non-linear
scale dependent bias. In the next section we develop these equations
further.


\section{The  Bispectrum Monopole}\label{sec:IsoBispec}

\subsection{Euler angle averages}

The redshift space bispectrum is an anisotropic function on the sphere
that depends on 5 variables.  The first three are the triangle
configuration, and the other two specify its orientation with respect
to the z-axis.  However it is common practice to measure this quantity
averaged over all possible orientations of the coordinate frame. Thus
to compare with this isotropized observable, we shall now derive the
isotropized form for the halo model, which we shall denote
$\overline{B}^s$. Interestingly, the following approach is identical
to that which one performs for the triaxial halo model in real space
\cite{SmithWatts2005,SmithWattsSheth2006}, since on small scales, one
may effectively think of transforming to redshift space as simply
transforming a set of spherical haloes into a set of prolate
ellipsoids whose semi-major axes all point along the line-of-sight.
Of course, the real situation is more complex, since the halo centers
are also distorted according to the halo velocity projected along the
line-of-sight, but nevertheless we may borrow some mathematical
machinery from the triaxial halo analysis.

The isotropic function is thus,
\ba \overline{B}^s(k_1,k_2,\theta_{12}) & = & \frac{1}{8\pi^2}\int
d\gamma_1\, d(\cos\gamma_2)\, d\gamma_3\, \nonumber \\ 
& &  \hspace{-1.4cm}\times\ B^s\left[\RotII\bk_1,\RotII\bk_{2}\right] \ ,\ea
where $\RotII$ is the rotation matrix for the components of the basis
vectors.  $\RotII$ is parametrized by three position or Euler angles,
($\gamma_1$,$\gamma_2$,$\gamma_3$) and these are the $z'-y'-z''$
rotation angles (see Appendix \ref{app:rotation} for the explicit form
of the matrix we use). Note that we assign uniform probability on the
sphere to each triple of angles.

The following considerations simplify this expression considerably.
Firstly, in the PT expressions for $B^s$, each term depends on either
the angle between two k-vectors or the projection of each vector along
the $z$-axis.  In the first case, the use of matrix notation shows
that
\ba \left[\bk_i^{\prime}\right]^T \bk_j^{\prime} & = & \left[\RotII
\bk_i\right]^T\RotII\bk_j \nonumber \\
& = & \bk_i^T\RotII^T\RotII\bk_j \nonumber \\
& = & \bk_i^T\bk_j \ ;\ea
this is the well-known result that scalar products are invariant under
rotations of the coordinate basis functions.  In the second case, the
projection of each rotated vector onto the line of sight direction (or
z-axis) can be written
\ba 
\bk^{\prime}\cdot\hat{\bf z} & = & \RotII\bk\cdot\hat{\bf z} \nonumber \\
& = & k_x \SgII\,\CgI  + k_y\SgII\,\SgI +k_z \CgII  \nonumber \\
& \equiv & A(\bk,\gamma_1,\gamma_2)\ ,\label{eq:rotatek}\ea
Thus, we see that the resultant function must be invariant under the
$\gamma_3$ rotation, and hence this integral may be computed
trivially.  Finally, because our final quantity $\overline{B}$ must be
independent of the initial locations of the $k$-vector triple, we may
without loss of generality choose these locations to be as convenient
as possible.  Therefore, we let the initial $\bk_1$ vector lie along
the polar-axis and constrain $\bk_2$ and $\bk_3$ to lie in the
$z$--$y$ plane: i.e.
\ba
\left[\bk_1\right]^T & = & \left( 0,\,0, k_1 \right) \nonumber \\
\left[\bk_2\right]^T & = & \left( 0,\, k_2 \cos\theta_{12},\, k_2\sin\theta_{12} \right)\nonumber \\
\left[\bk_3\right]^T & = & \left( 0,\, -k_2\cos\theta_{12},\,-k_1-k_2\sin\theta_{12}\right) 
\ea
where the last equality uses the closure condition: $\sum_i \bk_i =
0$. Thus, the $k$-vectors rotated into the new basis and dotted with
the $z$-direction are now written:
\ba
A_1 & = & \CgII (k_1)_z \ ;\nonumber \\
A_2 & = & \SgII\,\SgI (k_2)_y+\CgII (k_2)_z \ ;\nonumber \\
A_3 & = & -\SgII\,\SgI (k_2)_y-\CgII
[(k_1)_z+(k_2)_z] \ . \ \ \ \ \ea
The parameters $A_i\equiv A(\bk_i,\gamma_1,\gamma_2)$ are simply
related to the cosines of the $k$-vectors along the $z$-axis:
\be \mu_1 = \frac{A_1}{k_1}\ ;\ \ \ 
    \mu_2 = \frac{A_2}{k_2}\ ;\ \ \
    \mu_3 = -\mu_1\frac{k_1}{k_3}-\mu_2\frac{k_2}{k_3} ,\ee
where $q\equiv k_2/k_1$.  

We may now apply the above operation directly to our expressions for
the anisotropic bispectrum (Eqs~\ref{eq:bi1H}--\ref{eq:bi3H}). On
inserting the $A_i$ into each instance of $\mu_i$ in the density
profiles (Eq.~\ref{eq:redprofile}) and the halo center power spectra
and bispectrum (Eqs~\ref{eq:RedPowHaloCent} \&
\ref{eq:RedBiHaloCent2}), we find that the 1-, 2- and 3-Halo terms for
the bispectrum monopole become
\begin{widetext}
\ba 
\overline{B}^s_{\1H}(k_1,k_2,\theta_{12}) & = &
\frac{1}{4\pi\rhob_{\alpha}^3} \int d\gamma_1 \, d(\cos\gamma_2)\,\int
dM\, [W_{\alpha}]^3\,n(M)\, \prod_{i=1}^{3}\left\{U(k_i|M){\mathcal
V}(\mu_ik_i|M)\right\} \label{eq:RedBiIso1H}\ ;
\ea
\ba
\overline{B}^s_{\2H}(k_1,k_2,\theta_{12}) & = & \frac{1}{4\pi\rhob_{\alpha}^3}
 \int d\gamma_1\, d(\cos\gamma_2)\, 
  \!\!\prod_{i=\{1,2\}}\!\!
 \left\{ \int dM_i\, [W_{\alpha}]_i\, b_1(M_i) n(M_i)\,U(k_i|M_i)
{\mathcal V}(\mu_ik_i|M_i)\right\}
  \nonumber \\ & & \times [W_{\alpha}]_1\,U(k_3|M_1)\,{\mathcal V}(\mu_3k_3|M_1) 
  P_{11}(k_2)\left\{\frac{}{}1+\left[\beta_1+\beta_2\right]\mu_2^2+\beta_1\,\beta_2
\mu_2^4\right\}
+\cyc \ ;
  \label{eq:RedBiIso2H} 
\ea
\ba
\overline{B}^s_{\3H}(k_1,k_2,\theta_{12}) & = & \frac{1}{4\pi\rhob_{\alpha}^3}
 \int d\gamma_1\, d(\cos\gamma_2)\, \prod_{i=1}^3 
 \left\{ \int dM_i\, [W_{\alpha}]_i\,n(M_i)\,U(k_i|M_i)\,{\mathcal V}(\mu_ik_i|M_i)\right\}
  \nonumber \\ & & \times
  B^{s}_{\hc}(k_1,k_2,\theta_{12},\mu_1,\mu_2|M_1,M_2,M_3)
 \label{eq:RedBiIso3H}\ .
\ea
\end{widetext}
The only variables that depend on the Euler angles $\gamma_1$ and
$\gamma_2$ are $\mu_i$ and $A_i$. Each term requires evaluation of no
more than a 4-D embedded integral: two integrals for the Euler angles,
one for the mass, and one for the Fourier transform of the density
profile. Note that for simplicity, we have kept only the leading order
contribution to the the 2-Halo term. Technically this should be taken
up to the 1-Loop level to be consistent with the bispectrum which is
4th order in $\delta$. However, this issue is beyond the scope of the
current paper and will be addressed in \cite{Smithetal2008}.
%


\subsection{Computational considerations}

Our expressions for the bispectrum as presented above are complete.
However some calculational effort is still required before we may
attempt a practical implementation on the computer. We now present 
some simplifications.  

We begin by defining some convenient notation: 
Let $\psid$ and $\psiv$ denote the following integrals:
\ba 
\psidij & = & \frac{1}{\rhob_{\alpha}^j} \int dM  \,
n(M) \, b_i(M) \hspace{1cm} \nonumber \\ 
& & \hspace{-2cm}\times \,\prod_{l=1}^{j}\left\{\frac{}{}[W_{\alpha}]U(k_l|M)
{\mathcal V}(\mu_lk_l|M)\right\} \ ;\label{eq:psid}
\ea
\ba
\psivj & = & \frac{\fom}{\rhob_{\alpha}^j} \int dM  \,
n(M) \hspace{1cm}\nonumber \\ 
& & \hspace{-2cm}\times \,\prod_{l=1}^{j}\left\{\frac{}{}[W_{\alpha}] U(k_l|M)
{\mathcal V}(\mu_lk_l|M)\right\} \ .\label{eq:psiv} 
\ea
The first integral generalizes the halo bias weighting scheme applied
to the density field for the situation where $j$-points are within a
single halo.  The second integral generalizes the weighting scheme to
the similar situation for the halo velocity field. This notation has
some similarities with that of \cite{Seljak2001}, but is different in
the way in which the velocity field is treated -- recall that the
velocity field has been assumed to be unbiased.

In this notation we may re-write the 2- and 3-Halo terms in the
bispectrum; the 1-Halo term requires no simplification. Through
rearrangement of the mass integrals and expansion of the halo center
power spectrum through substitution of the appropriate kernels, we
find that the 2-Halo term can now be written
\begin{widetext}
\ba 
\overline{B}_{\alpha,\2H}^{s}(k_1,k_2,\theta_{12}) & = & \frac{1}{4\pi}
\int d\gamma_1\, d(\cos\gamma_2)\,
\left\{ P(k_2)\left[
\psi_{\delta,2}^{(1)}(\bk_1^{\prime},\bk_3^{\prime})\psi_{\delta,1}^{(1)}(\bk_2^{\prime}) 
\right. \right. \nonumber  \\
& & \hspace{-2cm}+
\left. \left. 
\left( \psi_{v,2}(\bk_1^{\prime},\bk_3^{\prime})\psi_{\delta,1}^{(1)}(\bk_2^{\prime})+
\psi_{\delta,2}^{(1)}(\bk_1^{\prime},\bk_3^{\prime})\psi_{v,1}(\bk_2^{\prime})\right) 
\mu_2^2 + \psi_{v,2}(\bk_1^{\prime},\bk_3^{\prime})
\psi_{v,1}(\bk_2^{\prime})\,\mu_2^4
\frac{}{}\right]+2\,\cyc \frac{}{}\right\}\ .\label{eq:BiRed2Hcompact}
\ea
\end{widetext}
The 3-Halo term is significantly more complex, owing to the halo
center bispectrum being the product of two first order kernels and one
second order kernel. Nevertheless, we may again isolate the integrals
over mass and use the $\psi$ functions to obtain 
\begin{widetext}
\ba 
\overline{B}^s_{\3H}(k_1,k_2,\theta_{12}) & = & \frac{1}{4\pi}
\int d\gamma_1\, d(\cos\gamma_2)\, 
\left\{\frac{}{}2P(k_1)P(k_2)T_1(\bk_1^{\prime},\bk_2^{\prime}) \right.
\nonumber\\
& & \hspace{-2cm}\times 
\left. \left[
T_2(\bk_1^{\prime},\bk_2^{\prime}|\bk_3^{\prime})\psi_{\delta,1}^{(1)}(\bk_3^{\prime})+
T_3(\bk_1^{\prime},\bk_2^{\prime}|\bk_3^{\prime})\psi_{v,1}(\bk_3^{\prime})+
{\mathcal W}_{12,3}\frac{\psi_{\delta,1}^{(2)}(\bk_3^{\prime})}{2}+   
T_4(\bk_1^{\prime},\bk_2^{\prime}|\bk_3^{\prime})\psi_{\delta,1}^{(0)}(\bk_3^{\prime})
\right] 
+2\, \rm cyc\right\}\ ; \ \ \label{eq:BiRed3Hcompact}
\ea
%
where we have defined the following useful quantities:
\ba 
T_1(\bk_i,\bk_j) & = & 
\prod_{m=i,j}
\left\{
\psi_{\delta,1}^{(1)}(\bk_m)+
\mu_m^2\left[\psi_{v,1}(\bk_m)+\fom\psi_{\delta,1}^{(0)}(\bk_m)\right]
\right\} \ ;\\
T_2(\bk_i,\bk_j|\bk_l) & = & 
F_2(\bk_i,\bk_j)+
\frac{1}{2}{\mathcal W}_{ij,l}\fom\mu_{ij}k_{ij} \left[\frac{\mu_i}{k_i}+\frac{\mu_j}{k_j}
\right] \ ;\\
T_3(\bk_i,\bk_j|\bk_l) & = & 
\mu_{ij}^2G_2(\bk_i,\bk_j)+
\frac{1}{2}{\mathcal W}_{ij,l}\fom\mu_{ij}k_{ij}
\left[\frac{\mu_i}{k_i}\mu_j^2+\frac{\mu_j}{k_j}\mu_i^2\right] \ ;\\
T_4(\bk_i,\bk_j|\bk_l) & = & 
\frac{1}{2}{\mathcal W}_{ij,l}\left[\fom\mu_{ij}k_{ij}\right]^2
\frac{\mu_i}{k_i}\frac{\mu_j}{k_j} \ ;
\ea
\end{widetext}
and   
\be {\mathcal W}_{ij,l}\equiv \frac{W(k_iR)W(k_jR)}{W(k_lR)} \ .\ee

The advantage of this reformulation of the 2-- and 3--Halo terms is
that we have decomposed the integrand into a set of algebraic
functions of the $\psi$ integrals (Eqs~\ref{eq:psid}--\ref{eq:psiv}),
and auxiliary functions, and these may all be computed in parallel
making the computation highly modular.


\subsection{The large-scale limit}\label{ssec:largescalelimit}

Next we consider the redshift space bispectrum in the very large scale
limit, as this should asymptotically reduce to the standard PT
expressions for the bispectrum, modulo discreteness corrections for
the point process associated with the halo field.  However, let us
first examine the large-scale limit of Eqs~(\ref{eq:psid})
and~(\ref{eq:psiv}). On letting $\left\{\bk_i\right\}\rightarrow 0$,
the density profile terms become $U^s(\bk_i)\rightarrow 1$, the window
function becomes ${\mathcal W}_{ij,l}\rightarrow 1$, and so
\ba
& & \lim_{\{k_j\}\rightarrow0} 
\psi_{\delta,j}^{(i)}=\frac{\left<b_i(M)[W_{\alpha}]^j\right>}
{\left<[W_{\alpha}]\right>^j} \ ;
\hspace{1cm} \nonumber \\
& & 
\lim_{\{k_j\}\rightarrow0}
\psi_{v,j}=\frac{\fom\left<[W_{\alpha}]^j\right>}{\left<[W_{\alpha}]\right>^j} \ ;
\label{eq:psilargescale}
\ea
where $\left<\dots\right>=\int dM p(M) \dots$, with $p(M)\equiv
n(M)/\nbarh$, $\nbarh$ being the total number density of haloes in the
required mass range. When $j=1$ we write these functions more simply as: 
the average non-linear bias parameter for the tracer particles, 
 $\psi_{\delta,1}^{(i)}\equiv {\overline{b}_{\alpha,i}}$, 
and the logarithmic growth factor for the linear velocity field 
 $\psi_{v,1}\equiv\fom$.

The bispectrum in the large-scale limit can now be computed directly. 
The 1-Halo term is trivially obtained, and the 2- and 3-Halo terms can 
be developed through replacing the $\psi$ functions in the general 
expressions (\ref{eq:BiRed2Hcompact}) and (\ref{eq:BiRed3Hcompact}), 
for their large scale forms (\ref{eq:psilargescale}).  After some 
algebraic manipulation we arrive at the result
\ba \overline{B}^s_{\alpha}(k_1,k_2,\theta_{12}) & = & 
\overline{B}^s_{\alpha,{\rm PT}}(k_1,k_2,\theta_{12}) \nonumber \\ 
& & \hspace{-2.0cm} + \frac{1}{\nbar_{\alpha,\2H}} \left[P(k_2)+P(k_3)+P(k_1)\right] 
+\frac{1}{\nbar_{\alpha,\1H}^2}\ ; \label{eq:bispecPTdisc}\ea
where the large scale PT bispectrum monopole in redshift space is
given by
\begin{widetext}
\ba 
\overline{B}^s_{\alpha,{\rm PT}}(k_1,k_2,\theta_{12}) & = &
\frac{1}{4\pi}\int d\gamma_1\, d(\cos\gamma_2) 
\prod_{i={1,2}}\left\{\overline{b}_{\alpha,1}
+\mu^2_i\fom\left[1+\overline{b}_{\alpha,0}\right]\right\}
\nonumber \\
& & \hspace{-3cm} \times \ 
\left\{
2 P(k_1) P(k_2) 
\left[\frac{}{}
\overline{b}_{\alpha,1} T_2(\bk_1,\bk_2|\bk_3)+
\fom T_3(\bk_1,\bk_2|\bk_3)+ \frac{\overline{b}_{\alpha,2}}{2}
+\overline{b}_{\alpha,0} \, T_4(\bk_1,\bk_2|\bk_3)
\right] 
\right\} + 2 \ \cyc \label{eq:RedPTBispec}
\ ;\ea
\end{widetext}
this is equivalent to that found by
\cite{Verdeetal1998,Scoccimarroetal1999}.  We also defined the 1-- and
2--Halo {\em `effective'} number densities to be:
\ba 
\frac{1}{\nbar_{\alpha,\1H}^2} & = & 
\frac{\left<W_{\alpha}^3\right>}{\left<W_{\alpha}\right>^3}\ ; \label{eq:effBiShot1H} 
\ea
\ba
\frac{1}{\nbar_{\alpha,\2H}} & = & 
\frac
{\left<b_1(M)[W_{\alpha}]^2\right>}{\left<W_{\alpha}\right>^2}
\overline{b}_{\alpha,1} 
+ \frac{\fom^2}{5}\, \frac{\left<[W_{\alpha}]^2\right>}{\left<W_{\alpha}\right>^3} \nonumber \\
& & \hspace{-1cm} + 
\frac{\fom}{3} \left[
\frac{\left<[W_{\alpha}]^2\right>}{\left<W_{\alpha}\right>^2} \overline{b}_{\alpha,1} 
+ 
\frac{\left<b_1(M)[W_{\alpha}]^2\right>}{\left<W_{\alpha}\right>^3} \right]  \ .
\label{eq:effBiShot2H}
\ea

Our final expression for the large-scale limit
(Eq.~\ref{eq:bispecPTdisc}) is similar to the standard theoretical
expectation for the bispectrum recovered from a Poisson point process
sampling of a continuous field (c.f. Section 43 in~\cite{Peebles1980}).
However, the halo model effective shot-noise terms (the last two terms
on the right hand side of Eq.~\ref{eq:bispecPTdisc}), are very
different from the standard form, which would simply have e.g.
$\nbar_{\alpha,\2H}=\nbar_{\alpha}$.  These effective number
densities, Eqs~(\ref{eq:effBiShot1H}) and (\ref{eq:effBiShot2H}),
represent the fact that in the halo model we assume that dark matter
haloes are Poisson sampled into the density field and that the tracer
particles are injected into these haloes, and so simply act as
different weights. The issue of sampling tracer particles into the
density field has some important implications for how one should
extract information from galaxy surveys. We shall reserve this
investigation for future work. 

In Appendix~\ref{app:shotnoise} we present a short discussion of how
these shot noise corrections can impact the reduced bispectrum.  The
main results are as follows: For standard shot noise, and in the
low-sampling limit there is no configuration dependence and $Q^{\rm
d}=1/3$, sub-script d denotes discrete. For the case where shot noise
is sub-dominant, the $Q^{\rm d}$ is reduced relative to the continuum
limit $Q$ for all configurations. In the halo model, and in the low
sampling limit, again there is no configuration dependence and $Q^{\rm
HM}=\left<[W_{\alpha}]^3\right>\left<[W_{\alpha}]\right>/3\left<[W_{\alpha}]^2\right>^2$.
For the case of sub-dominant shot noise, $Q^{\rm HM}$ is not
necessarily smaller than the continuum limit case. In
Sec.~\ref{sec:simulations} we show some tentative evidence for these
effects in the measurements from our numerical simulations.

Before continuing, it should also be noted that on setting $\fom=0$,
one may recover the 1-Loop PT bispectrum in real space for a set of
biased tracer particles $\alpha$.


\subsection{The small-scale limit and hierarchical models}\label{ssec:smallscalelimit}

On small scales, the bispectrum is dominated by the 1-Halo term, given
by Eq.~(\ref{eq:RedBiIso1H}). Our understanding of its behavior in
this limit can be guided by considering the case where all haloes are
of the same mass. In this situation we have
$n(M)\rightarrow\delta^D(M-M')\nbarh$ and
$\bar{\rho}_{\alpha}\rightarrow\nbarh
\left[W_{\alpha}\right]$. Applying these conditions to
Eq.~(\ref{eq:RedBiIso1H}), we have:
\ba \overline{B}^s_{\1H}(k_1,k_2,\theta_{12}|M) & = &
\frac{1}{4\pi\nbarh^2}\int d\gamma_1 \, d(\cos\gamma_2) \nonumber \\ &
& \hspace{-1.0cm}\times \, \prod_{i=1}^{3}\left\{U(k_i|M){\mathcal
V}(\mu_ik_i|M)\right\} \label{eq:RedBiIso1HMass}\ .  \ea
We may follow this same procedure for the power spectrum (see
Eq.~\ref{eq:RedPow1H}) and so construct the reduced bispectrum, whence
\be \overline{Q}_s=
\frac{ \int d\gamma_1 \, d(\cos\gamma_2)\,
\prod_{i=1}^{3}\left\{{\mathcal V}(\mu_ik_i|M)/U(k_i|M)\right\}/4\pi}
{ \mathcal{R}_{1,2}^{(0)}(a_1)\mathcal{R}_{1,2}^{(0)}(a_2)/U(k_3|M)^2 +
   2 \cyc}
\label{eq:1Hsmall}\ee
where $\mathcal{R}_{1,2}^{(0)}(a_1)$ is given by
Eq.~(\ref{eq:PowRed1HMonopole}). Notice that this expression no longer
depends on the weights $[W_{\alpha}]$ or number densities of tracers
$\nbarh$, but simply the real space profile and the 1-PT velocity
profile. If the density and velocity profiles were mass independent,
then Eq.~(\ref{eq:1Hsmall}) would predict the same configuration
dependence for all haloes. However, for realistic redshift space
profiles, this is not the case (see Appendix \ref{sec:details}). Thus,
if the halo model is a good description for small scale clustering,
then the hierarchical model is unlikely to be correct in real or
redshift space, and we may generally extend this statement to any
tracers of the density field. Finally, and somewhat interestingly,
notice that if the ratio ${\mathcal V}(\mu_ik_i|M)/U(k_i|M)$ is mass
independent, then the configuration dependence of the bispectrum
becomes universal, modulo an amplitude off-set.
 

\subsection{The White-Seljak Approximation}\label{ssec:ws}

The White-Seljak approximation (hereafter WS) is the supposition that
haloes are randomly oriented relative to each other so that in
computing the monopole of the bispectrum, orientation averages can be taken
separately over each individual halo and over the large scale
orientation of the halo relative to each other
\cite{White2001,Seljak2001}. In the triaxial halo model,
\cite{SmithWatts2005} showed that the overall contribution from halo
alignment to the matter power spectrum was negligible. Thus we may
similarly assume that on large-scales the 2-- and 3--Halo terms will
not be sensitive to the orientation of the FOGs -- and this allows us
to use the isotropic redshift space density profiles instead of the
anisotropic profiles. However, since the bispectrum is more sensitive
than the power spectrum to the shapes of structure, we shall be a
little more cautious, and demonstrate the validity of this
approximation in Section \ref{ssec:wsresults}.

In the WS approximation we therefore take,
\ba
& & \hspace{-0.7cm} \overline{\psi}_{\delta,1}^{(i)}(k_1) \equiv
\frac{1}{4\pi} \int d\gamma_1\, d(\cos\gamma_2)\, \psi_{\delta,1}^{(i)}(\bk_1^{\prime}) \ ,\nonumber \\
& & \hspace{-0.7cm} \overline{\psi}_{\delta,2}^{(i)}(k_1,k_2,\theta_{12}) \equiv 
\frac{1}{4\pi} \int d\gamma_1\, d(\cos\gamma_2)\, \psi_{\delta,2}^{(i)}(\bk_1^{\prime},\bk_2^{\prime})\ , 
\ea
and with similar expressions for the $\overline{\psi}_{v,j}$ functions. On
replacement of these terms into Eq.~(\ref{eq:BiRed2Hcompact}) we
find that the 2-Halo term simplifies to:
\ba \overline{B}_{\alpha,\2H}^{s,\WS}(k_1,k_2,\theta_{12}) 
& = & 
P(k_2) \left\{ \frac{}{} \overline{\psi}_{\delta,2}^{(1)}(k_1,k_3,\theta_{13})\overline{\psi}_{\delta,1}^{(1)}(k_2)
\right.  \nonumber \\
& & \hspace{-1.5cm} +
\frac{1}{3} \left[ \overline{\psi}_{v,2}(k_1,k_3,\theta_{13})\overline{\psi}_{\delta,1}^{(1)}(k_2)\right. \nonumber\\
& & \hspace{-1.5cm} +
\left. \overline{\psi}_{\delta,2}^{(1)}(k_1,k_3,\theta_{13})\overline{\psi}_{v,1}(k_2) \right]
\nonumber \\ 
& & \hspace{-1.5cm} +
\left. \frac{1}{5} \overline{\psi}_{v,2}(k_1,k_3,\theta_{13}) \overline{\psi}_{v,1}(k_2)
\frac{}{}\right\}+2\,\cyc  \ . \label{eq:BiRed2HcompactWS}
\ea
Similarly, on applying the WS approximation to
Eq.~(\ref{eq:BiRed3Hcompact}), the 3-Halo term reduces to:
\ba 
\overline{B}^{s,\WS}_{\alpha,\3H}(k_1,k_2,\theta_{12}) & = & 2 P(k_1) P(k_2) 
\int \frac{d\gamma_1}{4\pi}\, d(\cos\gamma_2)\, 
\nonumber \\
& & \hspace{-1.5cm} \times   \overline{T}_1(\bk_1^{\prime},\bk_2^{\prime}) 
\left[\frac{}{}
T_2(\bk_1^{\prime},\bk_2^{\prime}|\bk_3^{\prime})  \overline{\psi}_{\delta,1}^{(1)}(k_3) \right. \nonumber \\
& & \hspace{-1.5cm} +
T_3(\bk_1^{\prime},\bk_2^{\prime}|\bk_3^{\prime})  \overline{\psi}_{v,1}(k_3)  +
{\mathcal W}_{12,3}\frac{\overline{\psi}_{\delta,1}^{(2)}(k_3)}{2} \nonumber \\
& & \hspace{-1.5cm}+   
\left. \frac{}{}T_4(\bk_1^{\prime},\bk_2^{\prime}|\bk_3^{\prime})\overline{\psi}_{\delta,1}^{(0)}(k_3)
\right] +2\, \rm cyc \ ; \label{eq:BiRed3HcompactWS}
\ea
where 
\ba 
\overline{T}_1(\bk_i,\bk_j) & = & \prod_{m=i,j}
\left\{ \overline{\psi}_{\delta,1}^{(1)}(k_m) + \mu_m^2 \left[ \overline{\psi}_{v,1}(k_m) 
\right. \right. \nonumber \\
& & + \left.\left. \fom\overline{\psi}_{\delta,1}^{(0)}(k_m)\right]
\right\} \ . \ea

This completes our analytic investigation. In the next sections we
shall provide numerical evaluation of our expressions for the
bispectrum.


\begin{figure*}
\centerline{
\includegraphics[width=5.5cm]{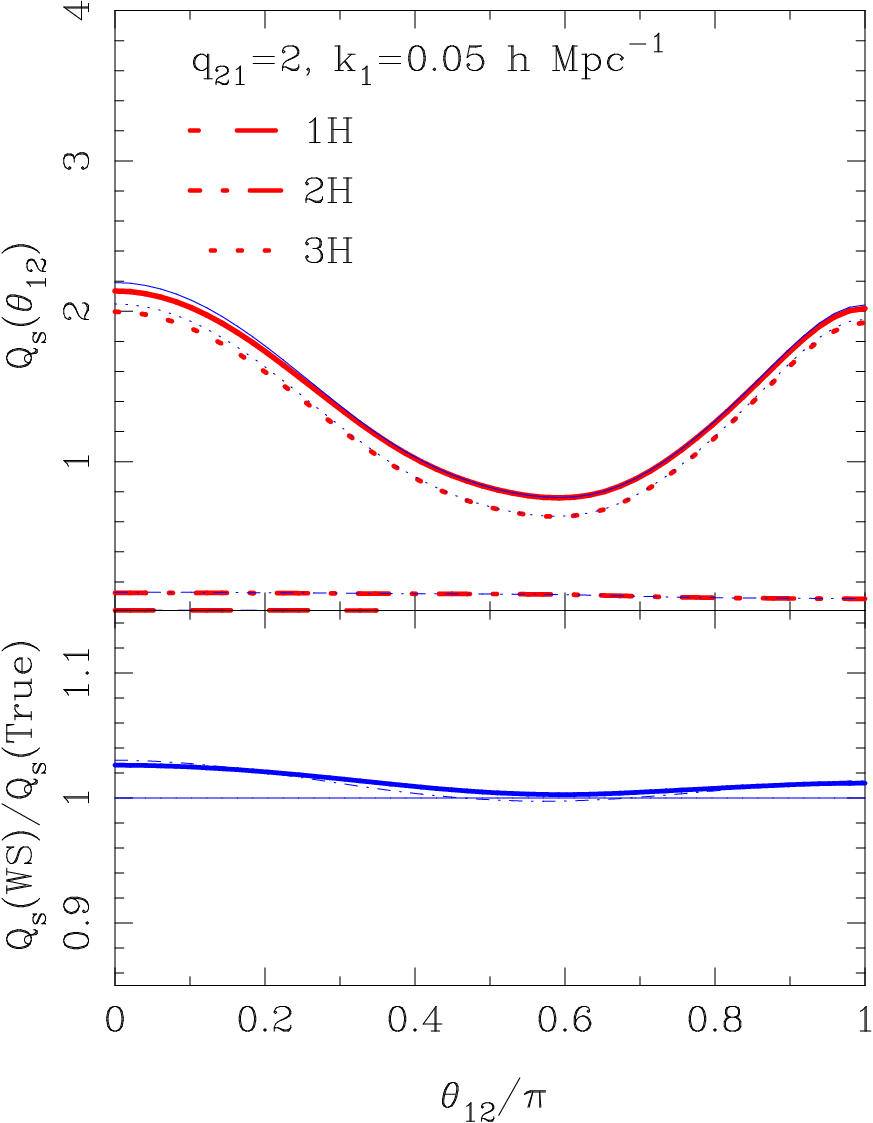}
\includegraphics[width=5.5cm]{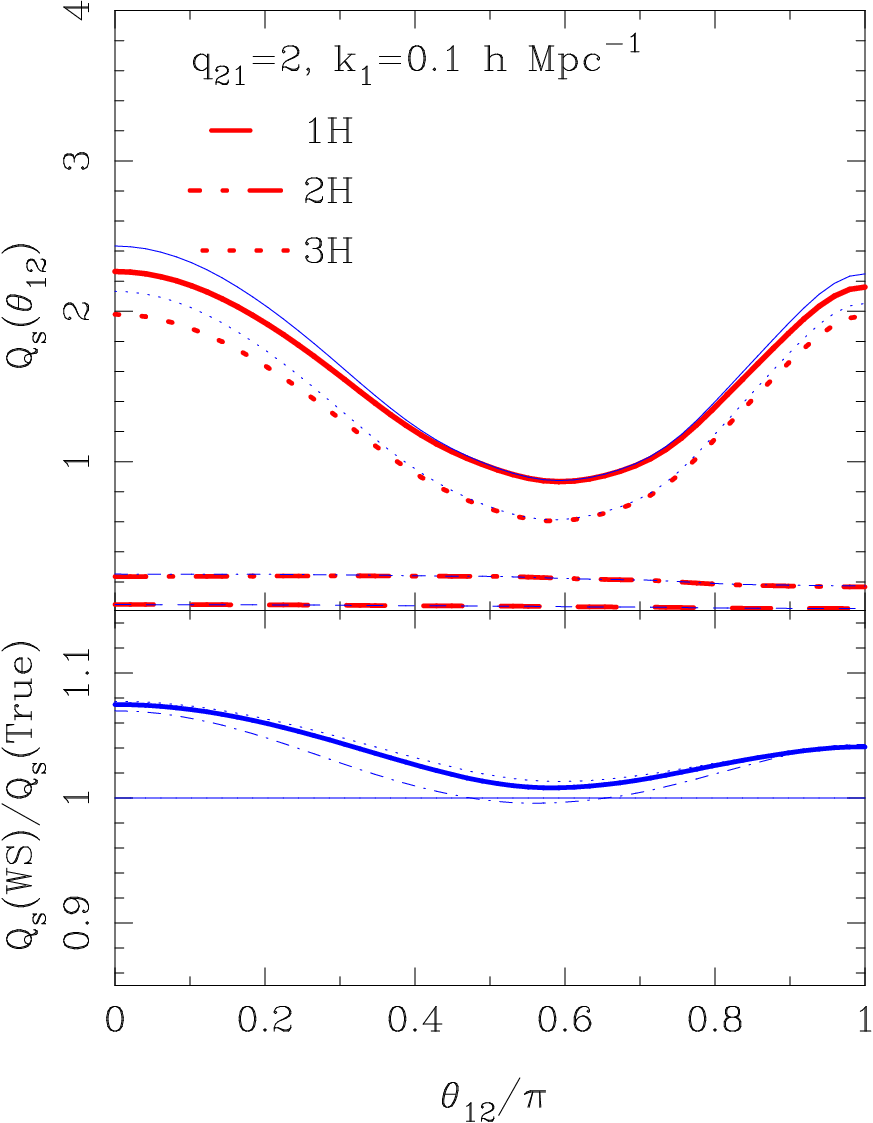}}
\centerline{
\includegraphics[width=5.5cm]{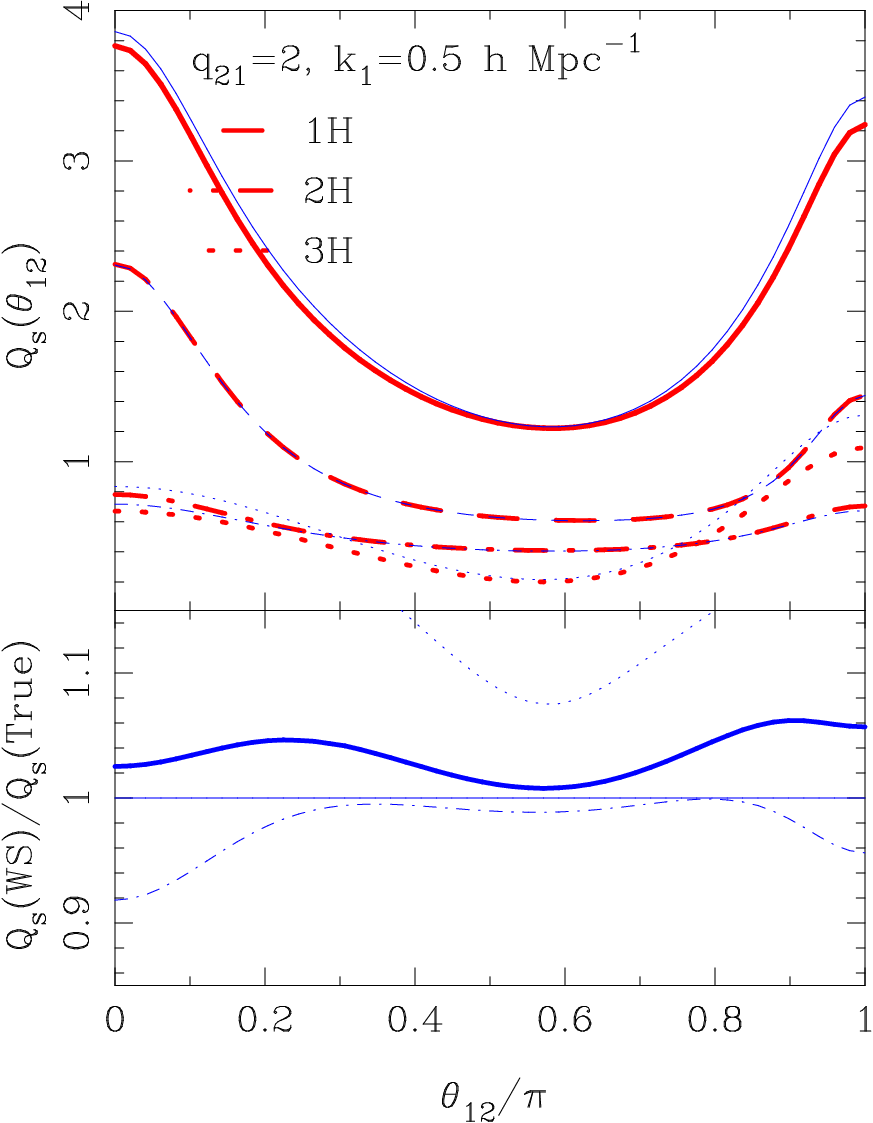}
\includegraphics[width=5.5cm]{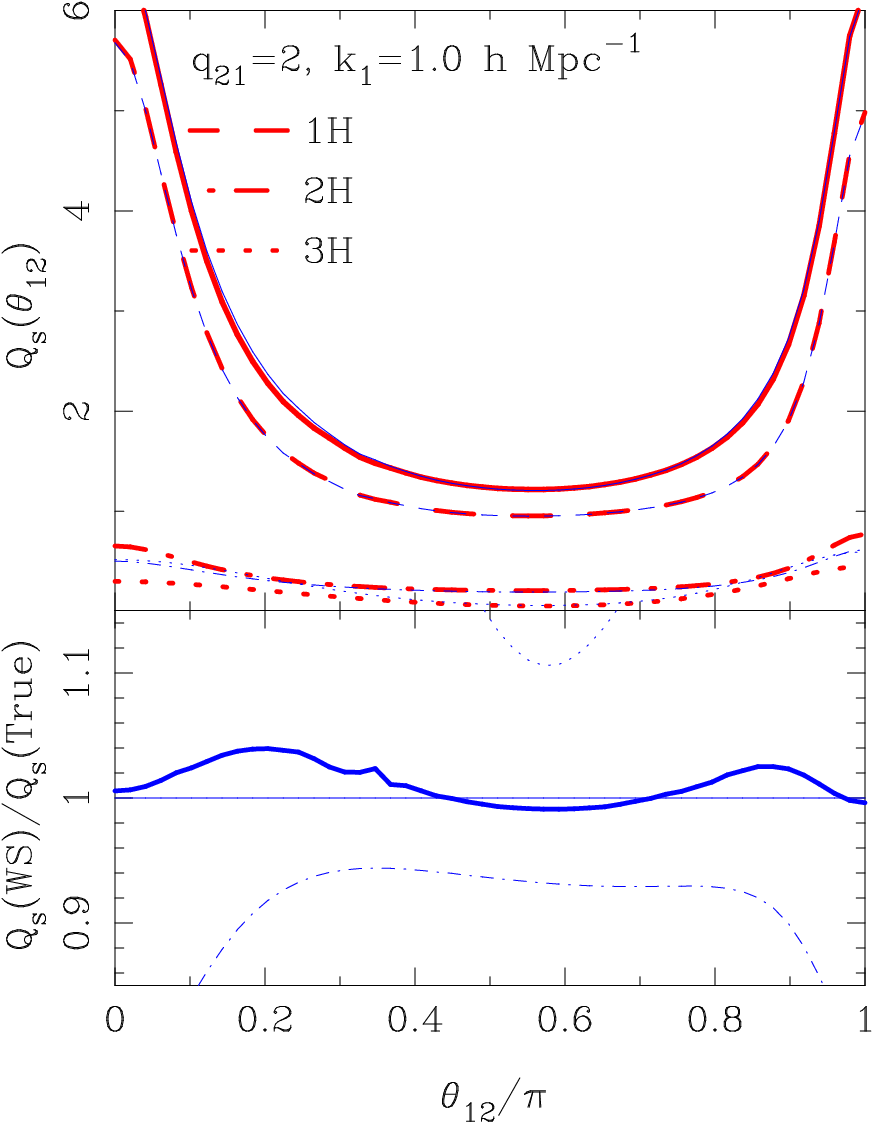}}
\caption{\small{Configuration dependence of the reduced bispectrum monopole in
 redshift space -- comparison of the predictions from the exact Halo
 Model expressions with those under the WS-approximation. Top sections
 of each panel show results for triangles with $k_2/k_1=2$ and on
 scales: $k_1=\{0.05,0.1,0.5,1.0\} h\,\Mpc^{-1}$. The dash, dot-dash
 and dotted lines in each panel correspond to the 1-, 2- and 3-Halo
 terms, respectively. The solid lines correspond to the sum. Thick
 (red) and thin (blue) lines are the exact Halo Model and WS
 approximate results, respectively. Bottom sections of each panel show
 the ratio of the total WS approximate bispectra to the exact Halo
 Model calculation. Line styles have same meaning as in top panels.}
\label{fig:QthetaRedWS}}
\end{figure*}


\begin{figure*}
\centerline{
\includegraphics[width=5.5cm]{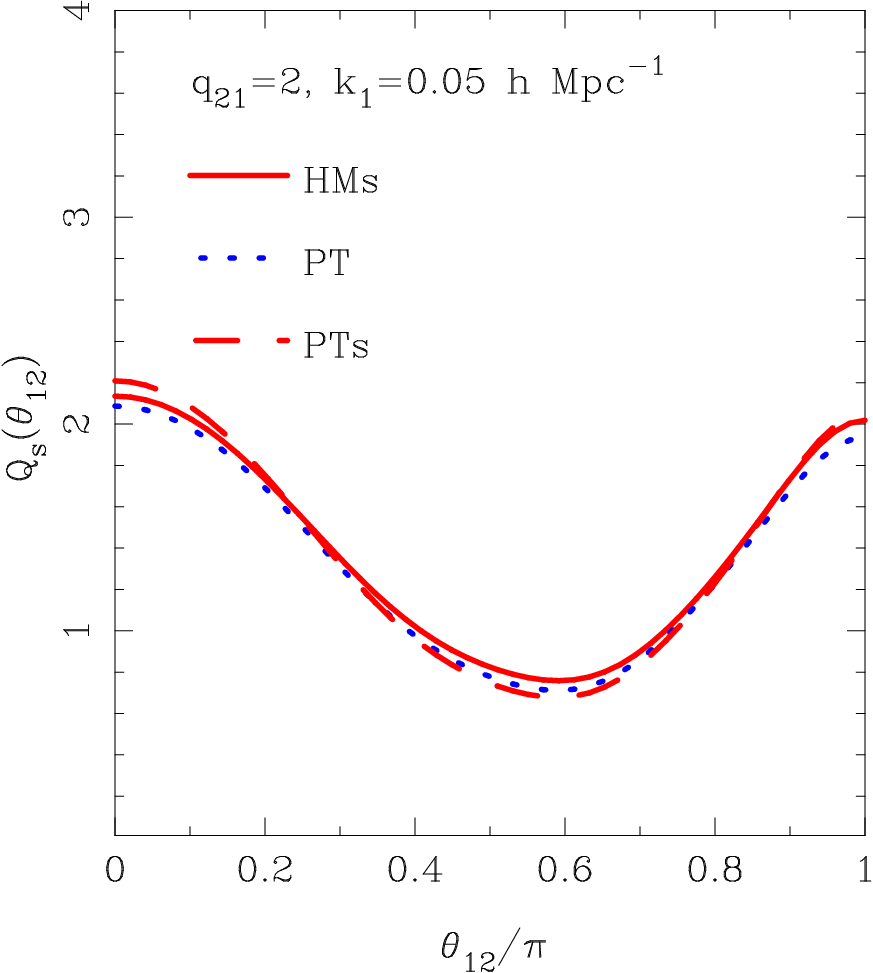}
\includegraphics[width=5.5cm]{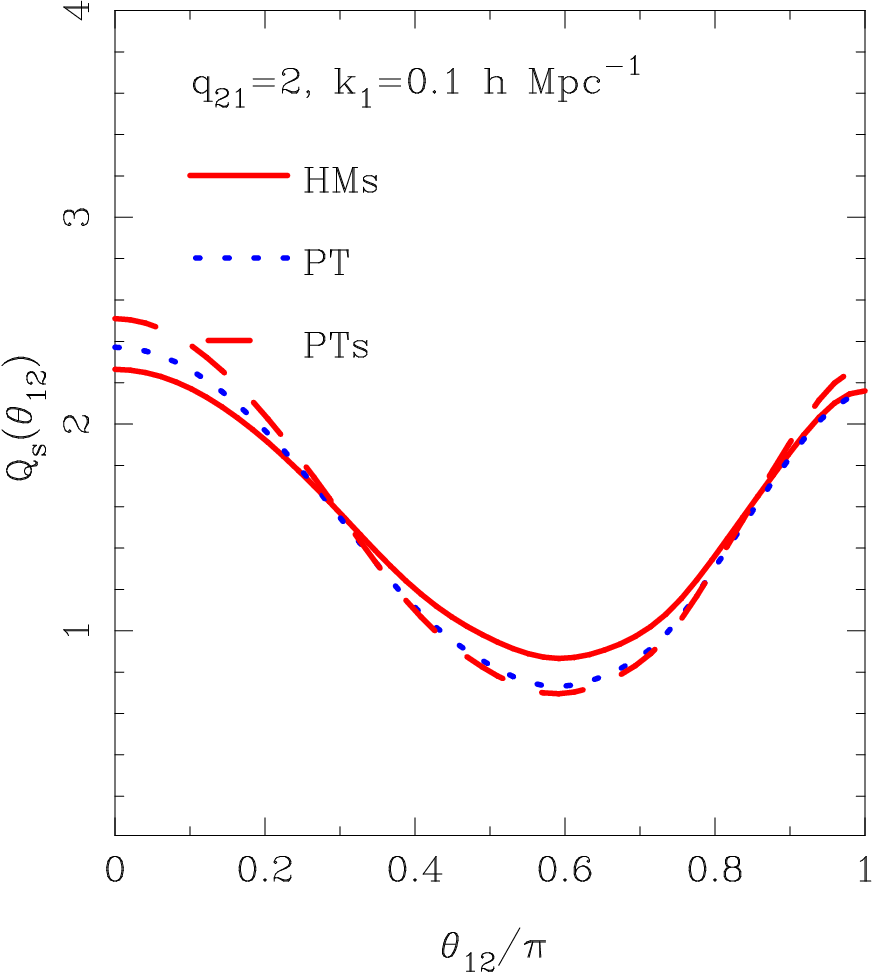}}
\centerline{
\includegraphics[width=5.5cm]{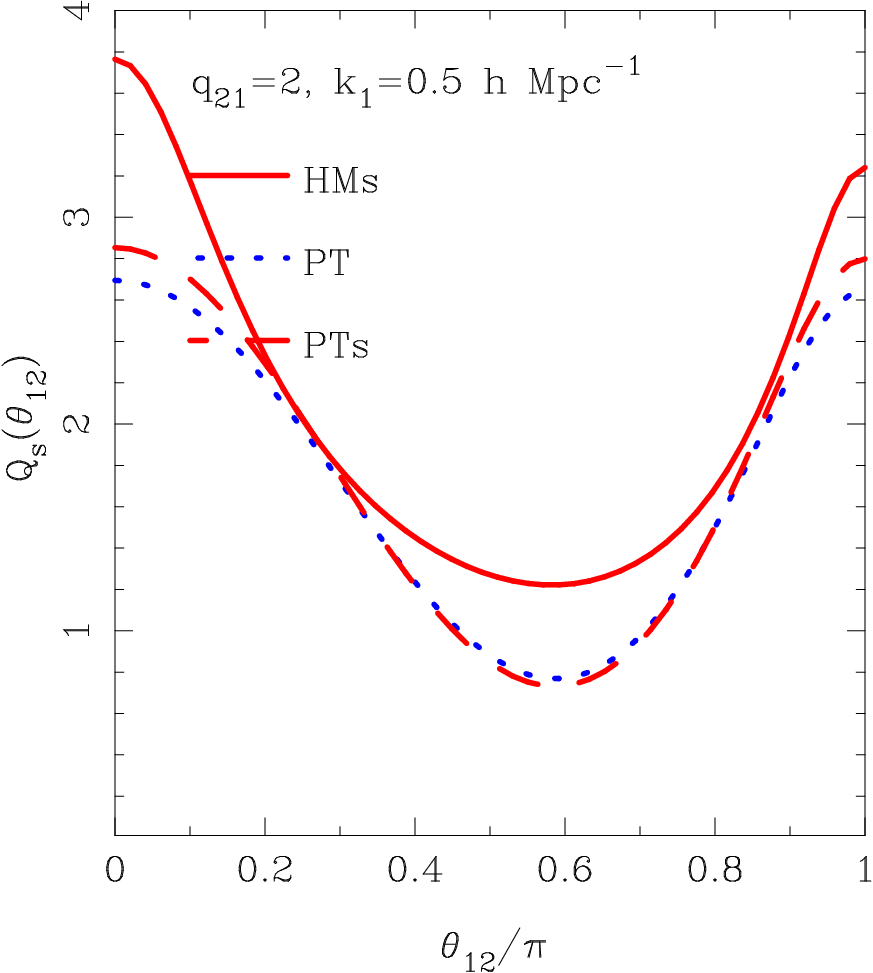}
\includegraphics[width=5.5cm]{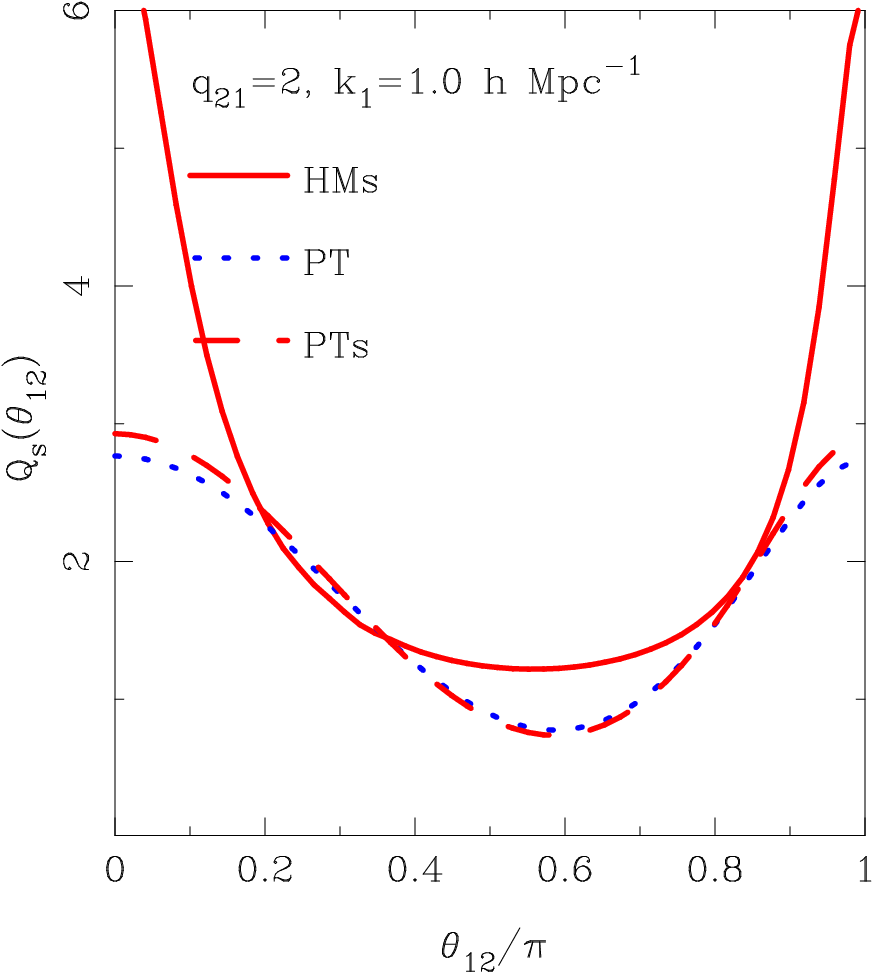}}
\caption{\small{Comparison of analytic model predictions with results
 from PT.  The configuration dependence is shown for for triangles
 with $k_2/k_1=2$ and on scales: $k_1=\{0.05,0.1,0.5,1.0\}
 h\,\Mpc^{-1}$. In all panels: solid (red) lines represent redshift
 space halo model predictions (HMs); dash (red) lines correspond to
 redshift space tree-level PT (PTs) \cite{Scoccimarroetal1999}; and
 dotted (blue) lines correspond to real space tree-level PT.}
\label{fig:QthetaRed}}
\end{figure*}


\section{Analytic results}\label{sec:results}

\subsection{Testing the WS approximation}\label{ssec:wsresults}

Figure \ref{fig:QthetaRedWS} compares the predictions for the
reduced redshift space matter bispectrum (Eqs~(\ref{eq:RedBiIso1H}),
(\ref{eq:BiRed2Hcompact}), and (\ref{eq:BiRed3Hcompact})) with the WS
approximate expressions Eqs~(\ref{eq:RedBiIso1H}),
(\ref{eq:BiRed2HcompactWS}), (\ref{eq:BiRed3HcompactWS}).  The four
panels show $k$-space triangles with $k_2/k_1=2$ and with
$k_1=\left\{0.05,0.1,0.5,1.0\right\} h\,\Mpc^{-1}$.

On very large scales (top left panel), $k=0.05 h\,\Mpc^{-1}$, the
predictions are dominated by the 3-Halo term, and the approximate
expressions (thin blue lines) are in almost perfect agreement with the
exact Halo Model predictions (thick red lines), with $<2\%$ deviations
across different configurations.  This is to be expected following our
derivation of the bispectrum in the large-scale limit,
c.f. Eq.~(\ref{eq:bispecPTdisc}); i.e., there is no dependence on the
halo profiles and so the WS approximation does not play a role here.  On
slightly smaller scales (top right panel) $k_1=0.1 h\,\Mpc^{-1}$,
small but significant departures are found at the level of $<10\%$. It
can be seen that these are entirely due to deviations in the 3-Halo
term -- the order in which we spherically average the profiles is now
important. However, on still smaller scales (bottom left panel)
$k_1=0.1 h\,\Mpc^{-1}$, whilst the discrepancy between the approximate
and exact 3-Halo term becomes larger, the 1- and 2-Halo terms begin to
dominate and so the difference in the total appears $<6\%$. Finally,
on the smallest scales considered, the 1-Halo term comes to fully
dominate and since no approximation is made here the results are in
good agreement, $<5\%$.

For a given scale, the largest deviations typically appear for the
case of colinear triangles, i.e. where all three $k$-vectors are
co-linear. This leads us to suppose that the equilateral bispectrum as
a function of scale will show agreement at the level of $<5\%$ across
a wide range of scales under this approximation. Since current
observational measurements of the bispectrum on large scales have
sample and cosmic variance errors of the order $\sim50\%$ on scales
$k_1\sim 0.1 h\,\Mpc^{-1}$, going down to several percent at $k_1\sim
1 h\,\Mpc^{-1}$, we anticipate that, at least for current data, the WS
approximation should be useful.  We highlight again that the main
advantage of this approximation is the increased speed with which one
can compute the bispectrum: the $\psid$ and $\psiv$ functions are only
evaluated once for a particular configuration, as opposed to thousands
of times. However, in all that follows we shall only show results for
the exact evaluation (no WS approximation) of our redshift space Halo Model.


\subsection{Comparison with perturbation theory}

Figure \ref{fig:QthetaRed} compares the analytic predictions for the
reduced bispectrum from our model with corresponding results from
PT. The four panels show again $k$-space triangles with with
$k_2/k_1=2$ and for the same scales as presented in
Fig.~\ref{fig:QthetaRedWS}. The solid (red) lines in each panel show
our HMs predictions (recall HMs means Halo Model in redshift space) in
the WS approximation. The (red) dash lines show the PTs predictions,
as given by our Eq.~(\ref{eq:RedPTBispec}). The (blue) dotted lines
correspond to real space PT predictions.
 
On the largest scales $k_1=0.05~h\,\Mpc^{-1}$ (top-left panel), the
HMs and PT results match almost perfectly: the configuration
dependence shows excess signal for colinear triangles, indicating
that on these very large scales non-linearity induces structures that
are, on average, more filamentary than spherical
\cite{Scoccimarroetal1998,Scoccimarroetal1999} (in this diagram
spherical perturbations are best probed by isosceles triangles,
$\theta_{12}\sim 2\pi/3$). However, we notice that there is a small
deviation $<5\%$ for the situation where the $k_1$ and $k_2$ vectors
are aligned. This owes to the fact that the 1- and 2-Halo terms are
non-vanishing as $k\rightarrow 0$, and as discussed in Section
\ref{ssec:largescalelimit}, this gives rise to an `effective
shot-noise' like behavior. A discussion of why co-linear $k_1$--$k_2$
triangles (i.e. $\theta_{12}=0$) are more preferentially affected is
given in Appendix \ref{app:shotnoise}). It should also be noted that
unlike for the case of standard shot noise, $Q^{\rm HM}>Q^{\rm PT}$
for configurations close to isosceles triangles.

On slightly smaller scales $k_1=0.1~h\,\Mpc^{-1}$, both the PT and PTs
predictions show a small increase in configuration dependence, with
the PTs having slightly more signal for colinear triangles than
PT. The HMs predictions are in qualitative agreement with PT. However,
the flattening off seen in the previous panel is now much more
apparent. Recalling the corresponding panel in
Fig.~\ref{fig:QthetaRedWS}, it can be seen that this is attributed to
the rapidly rising 1- and 2-Halo terms.  

On smaller scales still $k_1=0.5~h\,\Mpc^{-1}$, the PT and PTs
predictions continue their previous trends, exhibiting a slightly
increased configuration dependence. However, the HMs predictions are
quite different, having a very strong $U$-shaped configuration
dependence. This owes to the 1- and 2-Halo terms becoming dominant.

On the smallest scales considered $k_1=1.0~h\,\Mpc^{-1}$, the HMs
predictions show a dramatic configuration dependence, with a very
strong signal for colinear triangles and a broad plateau for
triangle shapes around isosceles configurations -- this is the `{\em
U}-shape', which was first noted in the bispectrum by
\cite{Scoccimarroetal1999} and in the 3-point correlation function by
\cite{GaztanagaScoccimarro2005}.  Recalling Fig.~\ref{fig:QthetaRedWS}
(bottom left panel), we see here that the prediction is completely
dominated by the 1-Halo term, and so this {\em U}-shape feature is
simply an imprint of the halo shape in redshift space -- the FOG.
This result constitutes a more direct demonstration of the discussion
from Section \ref{ssec:smallscalelimit}, that hierarchical models are
unlikely to be a good description for the higher order clustering
statistics.

In this section we have shown from purely theoretical considerations,
that to use the galaxy bispectrum on large scales
$k_1<0.1\,h\,\Mpc^{-1}$ as a precise tool for cosmology, one needs to
understand exactly how to include the FOG effect into the modeling
and also how to include non-trivial discreteness effects of matter. In
the next section we confront the model with results from numerical
simulations.


\section{Comparison with numerical simulations}\label{sec:simulations}


\subsection{Numerical simulations}

In order to test our redshift space bispectrum we generated an
ensemble of 8 LCDM simulations, these were identical in every way,
except that for each simulation different random realizations of the
initial Fourier modes were drawn. The cosmological parameters for the
ensemble were selected to be in broad agreement with the WMAP best fit
model \cite{Spergeletal2007}: $\Omega_m=0.27$, $\Omega_\Lambda=0.73$,
$\Omega_b=0.046$, $h=0.72$ and $\sigma_8(z=0)=0.9$.  We used the {\tt
cmbfast} \cite{SeljakZaldarriaga1996} code to generate the linear
theory transfer function, and we adopted the standard parameter
choices, and took the transfer function output redshift to be at
$z=0$. The initial conditions for each simulation were then laid down
at $z=49$ using the publicly available {\tt 2LPT} initial conditions
generator \cite{Scoccimarro1998,Crocceetal2006}. Subsequent
gravitational evolution of the equations of motion was then performed
using the publicly available {\tt Gadget2} code \cite{Springel2005}.
Each simulation was run with $N=400^3$ particles in a comoving volume
of length $L=512^3 \Mpc h^{-1}$ and with a comoving force softening
set to 70 kpc$/h$. The simulations were performed using a 4-way dual
core Opteron processor system, and each ran to completion in roughly
$\sim$1800 timesteps from redshift $z=49$ to $z=0$ and this translates
to $\sim$2 days of wall clock time.


\subsection{Estimating the power spectrum}\label{ssec:estpower}

Before comparing our analytic model with the bispectra estimates from
the simulations, it is instructive to compute the real and redshift
space power spectra of the $z=0$ outputs of the ensemble. 

The density Fourier modes are estimated using the conventional Fast
Fourier Transform (FFT) method: the dark matter particles were
assigned to a regular cubical grid using the `Cloud-In-Cell' (CIC)
scheme \cite{HockneyEastwood1981}. The FFT of the gridded density
field was then computed using the publicly available {\tt FFTW}
routines \cite{FFTW}. Each resulting Fourier mode was then corrected
for the convolution with the mesh by dividing out the Fourier
transform of the mass-assignment window function. For the CIC
algorithm this corresponds to the following operation:
\be \delta_{\rm d}(\bk)=\delta_{\rm g}(\bk)/W_{\rm CIC}(\bk) \ ,\ee
where
\be W_{\rm CIC}(\bk)=\prod_{i=1,3}\left\{\left[\frac{\sin 
\left[\pi k_i/2k_{\rm Ny}\right]}{\left[\pi k_i/2k_{\rm Ny}\right]}\right]^2\right\} \ee
where sub-script d and g denote discrete and grid quantities, and where
$k_{\rm Ny}=\pi N_{\rm g}/L$ is the Nyquist frequency of the mesh and
$N_{\rm g}$ is the number of grid cells.

The power spectra of the discrete particles on scale $k_l$ are then
estimated by performing the following sums,
\be \widehat{\overline{P}}_{\rm
d}(k_l)=\frac{\Vu}{M}\sum_{l=1}^{M}\left|\delta_{\rm
d}(\bk_l)\right|^2 \, ,\label{eq:powestimate}\ee
where $M$ is the number of Fourier modes in a spherical shell in
$k$-space of thickness $\Delta k$. Note that the mode-by-mode
correction differs from the analysis of \cite{Smithetal2003,Jing2005}
where the correction for charge assignment was performed by computing
the spherically averaged window and dividing it out. Mode-by-mode
correction for the power spectrum was also performed in
\cite{Scoccimarroetal1998,SmithScoccimarroSheth2007}.


\begin{figure}
\centerline{ \includegraphics[width=8.5cm]{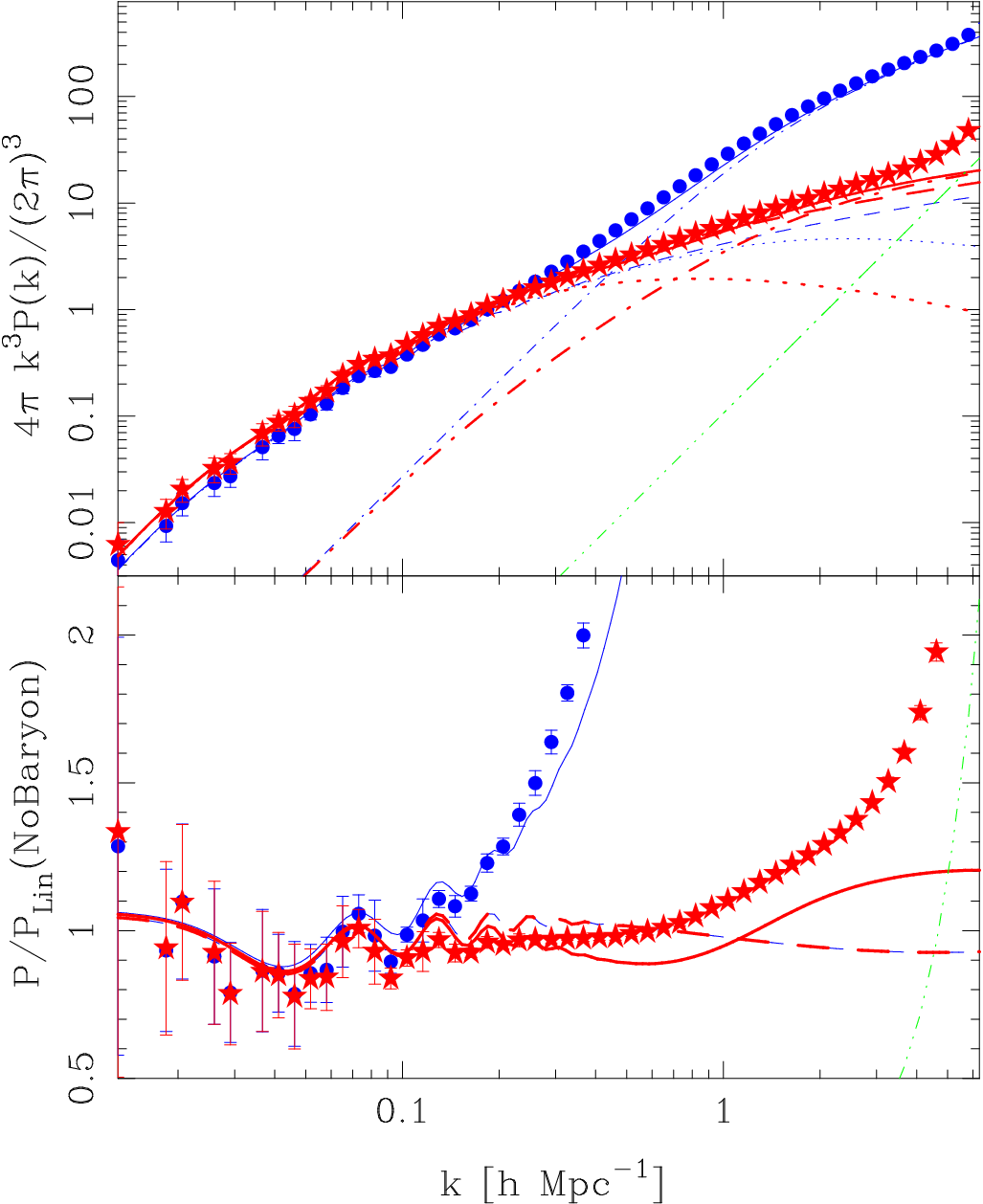}}
\caption{\small{Real and redshift space power spectrum of dark matter
particles at $z=0$ measured from the ensemble of LCDM simulations.
Top panel: The power spectrum. Blue and red points show real and
redshift space quantities, respectively. The solid, dot-dash and dotted
lines show the total halo model, 1-Halo and 2-Halo terms.  The triple
dot-dash curve shows the predictions for the Poisson shot noise
error. The dash line shows the linear theory. Bottom panel: The
measured power spectra ratioed with a no-baryon linear theory power
spectrum that has the same overall transfer function shape. Note that
for the redshift space spectra, we have scaled out the Kaiser boost.}
\label{fig:LCDMPowRealRed}}
\end{figure}


Figure \ref{fig:LCDMPowRealRed} shows the mean and 1-$\sigma$ errors
for the power spectra of dark matter particles in both real and
redshift space. The errors are computed directly from the 8
realizations. The spectra were computed using a $1024^3$ FFT and we
show all frequencies up to the Nyquist frequency -- the highest
$k$-modes show signs of increased power from both Poison shot noise
(triple dot-dash green line) and the aliasing of power from smaller
scales. We note that on the largest scales probed, the power spectra
show a sequence of wiggles, these are the well known Baryonic Acoustic
Oscillations (BAO) that have been discussed much in recent times
\cite{EisensteinHu1998,Meiksinetal1999,Eisensteinetal2005,SmithScoccimarroSheth2007,SmithScoccimarroSheth2008}
-- we shall not discuss these in this paper. The solid lines show the
total halo model predictions in real (blue lines) and redshift space
(red lines). We see that, whilst the real space model does reasonably
well, with an accuracy of the order $\sim10\%$, the redshift space
predictions fare less well, especially for scales $k>0.2
h\Mpc^{-1}$. Here the model systematically underpredicts the data by
roughly $\sim20\%$.

These predictions were very sensitive to how we modeled the FOG
effects, i.e.  the 1D velocity dispersion of particles within haloes
$\nu(k|M)$. Originally, we had simply used Eq.~(\ref{eq:sigma}) with
$\epsilon=1$, however in this case the predictions were particularly
poor.  We therefore investigated $M-\sigma_{\rm 1D}$ relation in our
simulations more closely.  Using a standard Friends-of-Friends (FoF)
algorithm with link length $f=0.2$, we located all haloes with
$M>1.0\times10^{13} M_{\odot}/h$. The 1D Velocity dispersions were
then estimated for each halo and binned as a function of mass. We
found that $\epsilon=1$ overpredicted the measured values by $>20\%$.
Fitting on $\epsilon$ it was found that $\epsilon=0.76$ provided a
better fit to the data, but it was not perfect, it having an accuracy
no better than $\sim10\%$ (see Fig.  \ref{fig:SigmaMass}). As can be
seen from the figure using $\epsilon=0.76$ does not generate the
correct power spectrum. 

A possible reason for this discrepancy could be that the 2-Halo term
in the redshift space power spectrum only includes linear halo motions
-- non-linear terms do contribute to this term (see
\cite{Shethetal2001,Scoccimarro2004,SmithScoccimarroSheth2008}) and
inclusion of these non-linear corrections may help alleviate this
problem. Another, is that the concentration--mass relation, which is
vital for getting the correct normalization of the halo density
profiles, may not be sufficiently accurate. Indeed we note a small
discrepancy between the real space measurements and Halo Model
predictions. Recent improvements on this relation by
\cite{Netoetal2007} may help to alleviate this problem. Also, we have
neglected the scatter in the concentration parameter, and it is well
known that this can change the predictions by a few tens of percent in
the nonlinear regime \cite{HuCooray2001}. Additionally, there is the
issue of halo triaxiality
\cite{SmithWatts2005,SmithWattsSheth2006}. We shall not pursue these
subtle corrections here, since our purpose is to simply present the
theoretical framework and show that it gives reasonable agreement with
the simulation data.


\subsection{Estimating the bispectrum}\label{ssec:estimation}

Our estimator for the bin and spherical averaged bispectrum was
developed following the work of \cite{Scoccimarroetal1998}, but with
some changes. Our estimator can be written:
\ba \widehat{\overline{B}}_{\rm d}(k_1,k_2,\theta_{12}) & = &
\frac{1}{V_1V_2}\int_{V_1,\,V_2}
\frac{\dq_1}{(2\pi)^3}\frac{\dq_2}{(2\pi)^3} \nonumber \\ & & \times \
\widehat{B}_{\rm d}(\bq_1,\bq_2,-\bq_1-\bq_2) \ ,\ea
where 
\be V_i=\int^{k_i+\Delta k/2}_{k_i-\Delta k/2} \frac{\dq}{(2\pi)^3} = 
\frac{4\pi k_i^2 \Delta k}{(2\pi)^3} 
\left[1+\frac{(\Delta k)^2}{12 k_i^2}\right] \ .
\ee
A practical implementation of this estimator involves computing the
following sum
\be \widehat{\overline{B}}_{\rm d} =\frac{\Vu^2}{N_{\rm tri}}
\sum_{{\bf n}_1,{\bf n}_2}^{N_{\rm tri}} {\mathcal Re}[\delta_{\rm
d}(\bk_{\bn_1})\delta_{\rm d}(\bk_{\bn_2})\delta_{\rm
d}(\bk_{-\bn_1-\bn_2})] \ , \label{eq:bispecEstimate}\ee
where ${\bf n}_i$ is an integer vector from the $k$-space origin to a
mesh point and so labels the modes, and where $N_{\rm tri}$ represents
the number of independent momentum conserving $k$-vector triangles in
the shells $V_1$ and $V_2$. In the above we take only the real part of
the product of the three Fourier modes, owing to the reality of the
bispectrum (see Eq.~\ref{eq:realspectra}). Note that when computing
the sums over $k$-space triangles we randomly sample modes from the
set of all possible triangles. Typically we limit the computations to
$10^4$ modes per shell (i.e. $10^8$ triangles). This method gives a
sufficient number of independent $k$-space triangles for a high
accuracy estimate, whilst keeping code execution times tolerable.

To accurately estimate $Q$ we are also required to estimate the
combination $\overline{Q}_{\rm fac}\equiv
\overline{P}(k_1)\overline{P}(k_2)+\overline{P}(k_2)\overline{P}(k_3)
+\overline{P}(k_3)\overline{P}(k_1)$. There are several approaches to
achieving this: one, we could simply estimate the power spectrum as in
Eq.~(\ref{eq:powestimate}) and then construct $\overline{Q}_{\rm fac}$
from this; alternatively one can compute an estimate of
$\overline{Q}_{\rm fac}$ using only those same modes that are used to
estimate $\overline{B}$.  We adopt this latter approach since expect
that it will reduce sample variance. Along with our estimates for $B$,
we also therefore accumulate
\ba \widehat{\overline{P}}_{i,\rm d} & = & \frac{\Vu}{N_{\rm tri}}
\sum_{{\bf n}_i}^{N_{\rm tri}} \left|\delta_{\rm
d}(\bk_{\bn_i})\right|^2 \ ;\ \ \ i\in\left\{1,2\right\}\ \ ;\ \\
\widehat{\overline{P}}_{3,\rm d}(\theta_{12}) & = & \frac{\Vu}{N_{\rm
tri}} \sum_{{\bf n}_3}^{N_{\rm tri}} \left|\delta_{\rm
d}(\bk_{\bn_3}[\theta_{12}])\right|^2 \ \ ;\\
\widehat{\overline{Q}}_{\rm fac,d}(\theta_{12}) & \equiv &
\widehat{\overline{P}}_{1,\rm d}\widehat{\overline{P}}_{2,\rm d}+
\widehat{\overline{P}}_{2,\rm d}\widehat{\overline{P}}_{3,\rm d}+
\widehat{\overline{P}}_{3,\rm d}\widehat{\overline{P}}_{1,\rm d}\ .
\ea
We draw close attention to the fact that these estimates for the power
spectra $\widehat{\overline{P}}_1$, $\widehat{\overline{P}}_2$ and
$\widehat{\overline{P}}_3$ are not the same as in
Eq.~(\ref{eq:powestimate}): in the above case we average over all
$k$-space triangles that are used and not just the unique modes. We
have also made it explicitly clear that the estimates for
$\widehat{\overline{P}}_1$ and $\widehat{\overline{P}}_2$ do not
change with $\theta_{12}$, but that $\widehat{\overline{P}}_3$ does.
Following this procedure helps to reduce cosmic variance.  We also
note the following pitfall: had we estimated $Q_{\rm fac}$ for each
$k$-space triangle and then averaged these estimates over all
triangles, i.e. taken $\overline{Q}_{\rm fac}\equiv
\overline{P(k_1)P(k_2)}+2\cyc$, then we would have been angle
averaging products of power spectra. In real space, where $P$ is an
isotropic function on the sphere, this makes no difference, however in
redshift space, where $P$ is anisotropic, this approach would be
incorrect and no U-shape would be seen in $Q$.

In order to correct $B$ and $P$ for discreteness we also estimate the
shot noise terms as \cite{Peebles1980}: 
\ba 
\widehat{\overline{P}}_{\rm shot} & \equiv & \frac{\Vu}{N} \ ;\\
\widehat{\overline{B}}_{\rm shot} & \equiv & 
\frac{\Vu}{N}\left[\widehat{\overline{P}}_{1,\rm d}
+\widehat{\overline{P}}_{2,\rm d}+
\widehat{\overline{P}}_{3,\rm d}\right]+\frac{\Vu^2}{N^2} \ ; \\
\widehat{\overline{Q}}_{\rm fac,\,shot} & \equiv & 2\frac{\Vu}{N}
\left[
\widehat{\overline{P}}_{1,\rm d}+
\widehat{\overline{P}}_{2,\rm d}+
\widehat{\overline{P}}_{3,\rm d}
\right]+3\frac{\Vu^2}{N^2}\ .\ \ \ \ \ 
\ea
Estimates of the shot noise corrected continuous spectra are then
arrived at through the following set of operations:
\ba 
\widehat{\overline{P}} & = & \widehat{\overline{P}}_{\rm d}-\widehat{\overline{P}}_{\rm shot} \ ;\\
\widehat{\overline{B}} & = & \widehat{\overline{B}}_{\rm d}-\widehat{\overline{B}}_{\rm shot} \ ;\\
\widehat{\overline{Q}}_{\rm fac} & = &
\widehat{\overline{Q}}_{\rm fac,\,d}-\widehat{\overline{Q}}_{\rm fac,\,shot} \ ;\\
\widehat{\overline{Q}} & = & \widehat{\overline{B}}/\widehat{\overline{Q}}_{\rm fac}\ .
\ea
In our measurements we shall show both spectra with and without the
shot noise corrections. 

Lastly, as a consistency check, we also estimate the imaginary
bispectrum, which is given by Eq.~(\ref{eq:bispecEstimate}) only now
we take the imaginary piece of the product. Thus, if our estimate is
correct, then this quantity should on average be zero. However, as the
number of independent triangles becomes small the imaginary piece may
become non-zero due to statistical fluctuations.


\begin{figure*}
\centerline{
\includegraphics[width=8.cm]{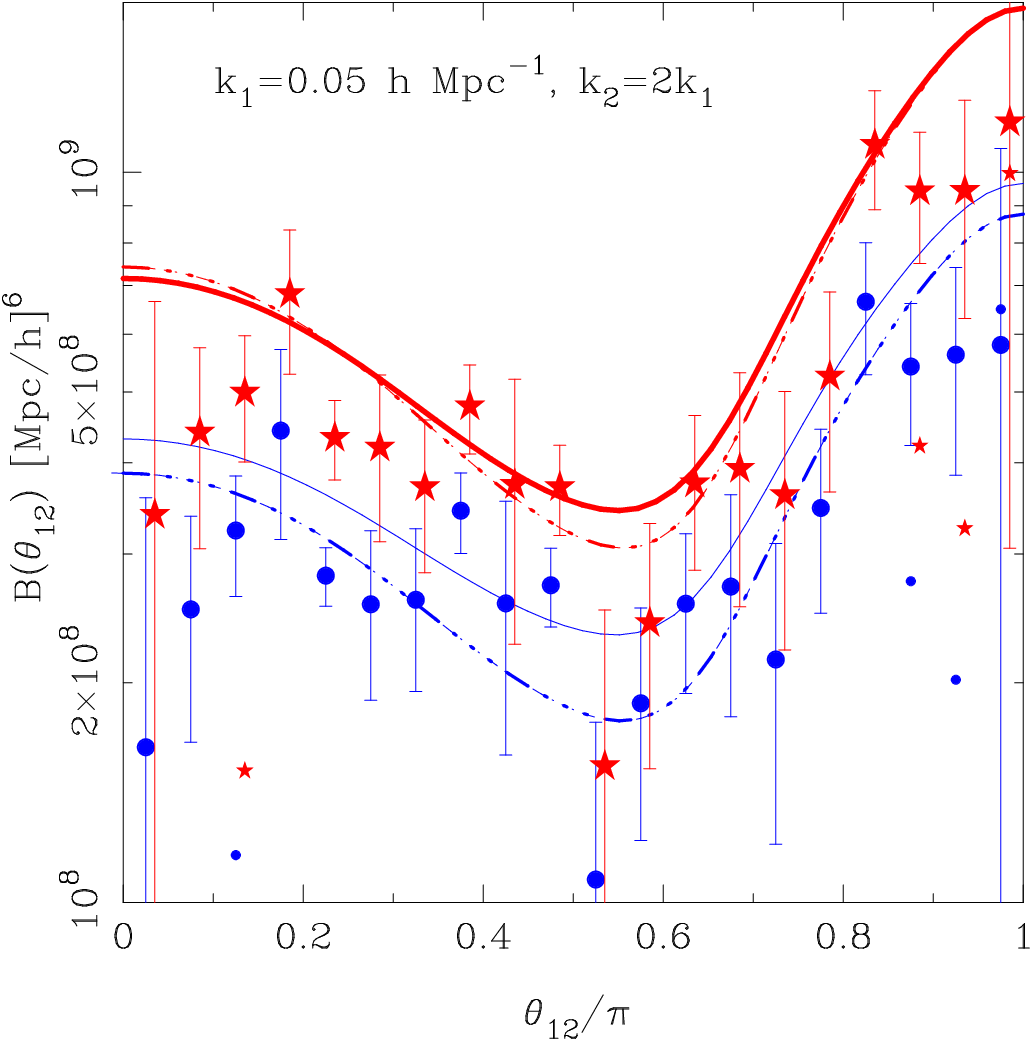}
\includegraphics[width=8.cm]{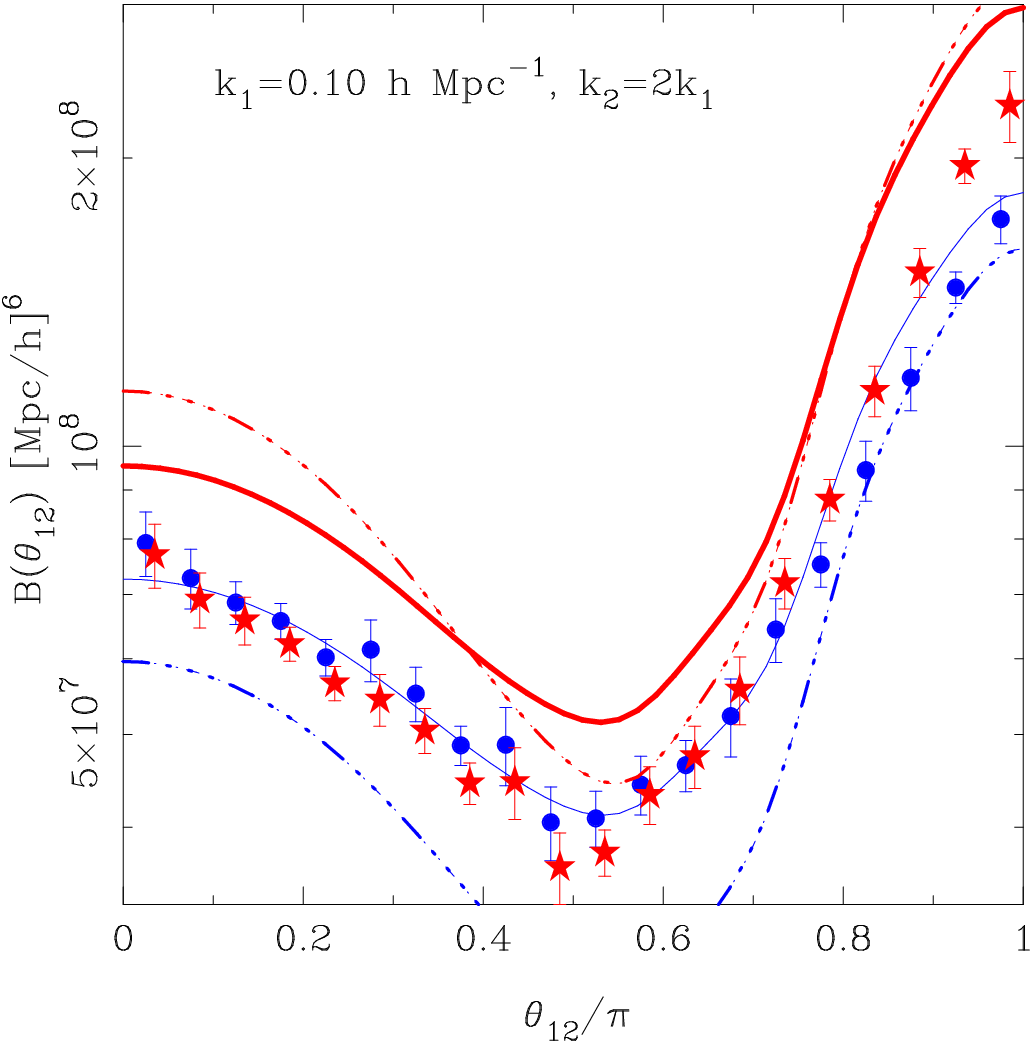}}
\centerline{
\includegraphics[width=8.cm]{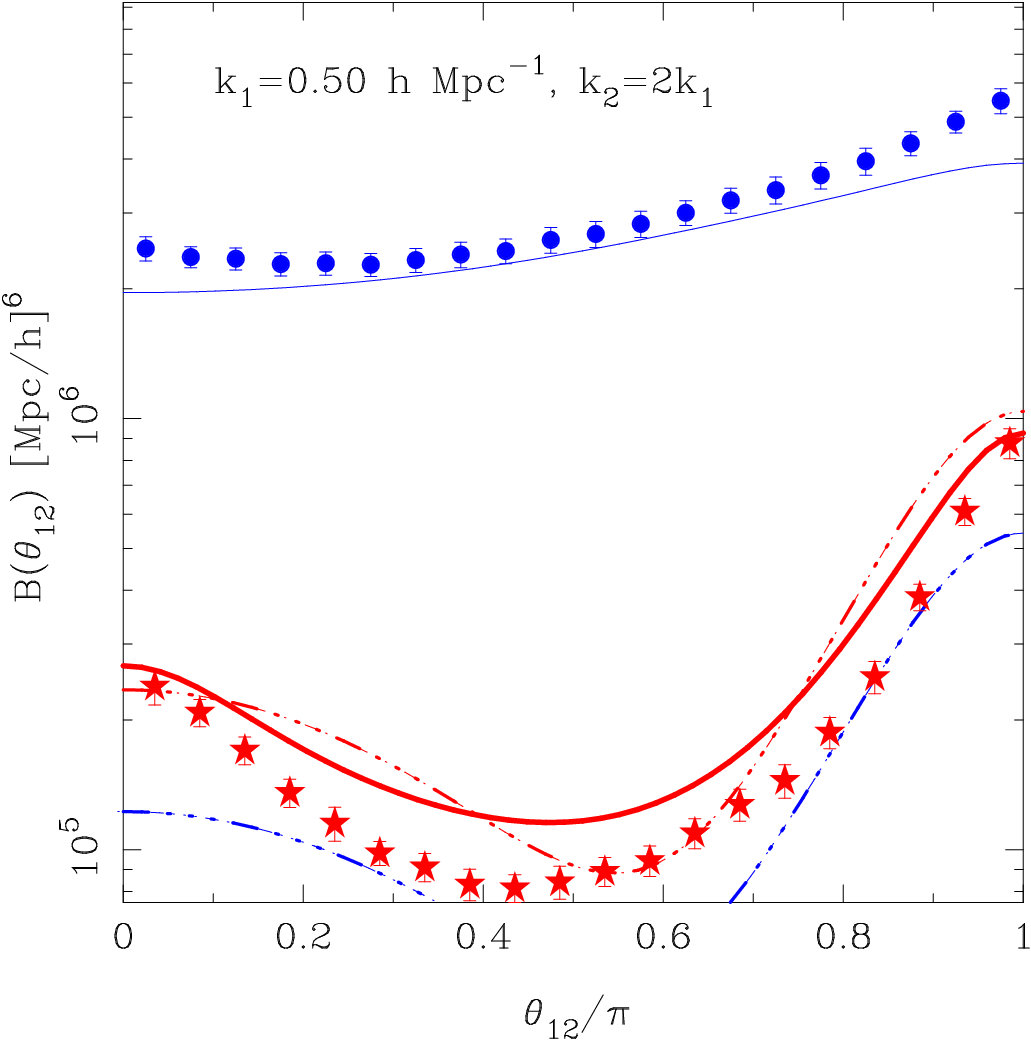}
\includegraphics[width=8.cm]{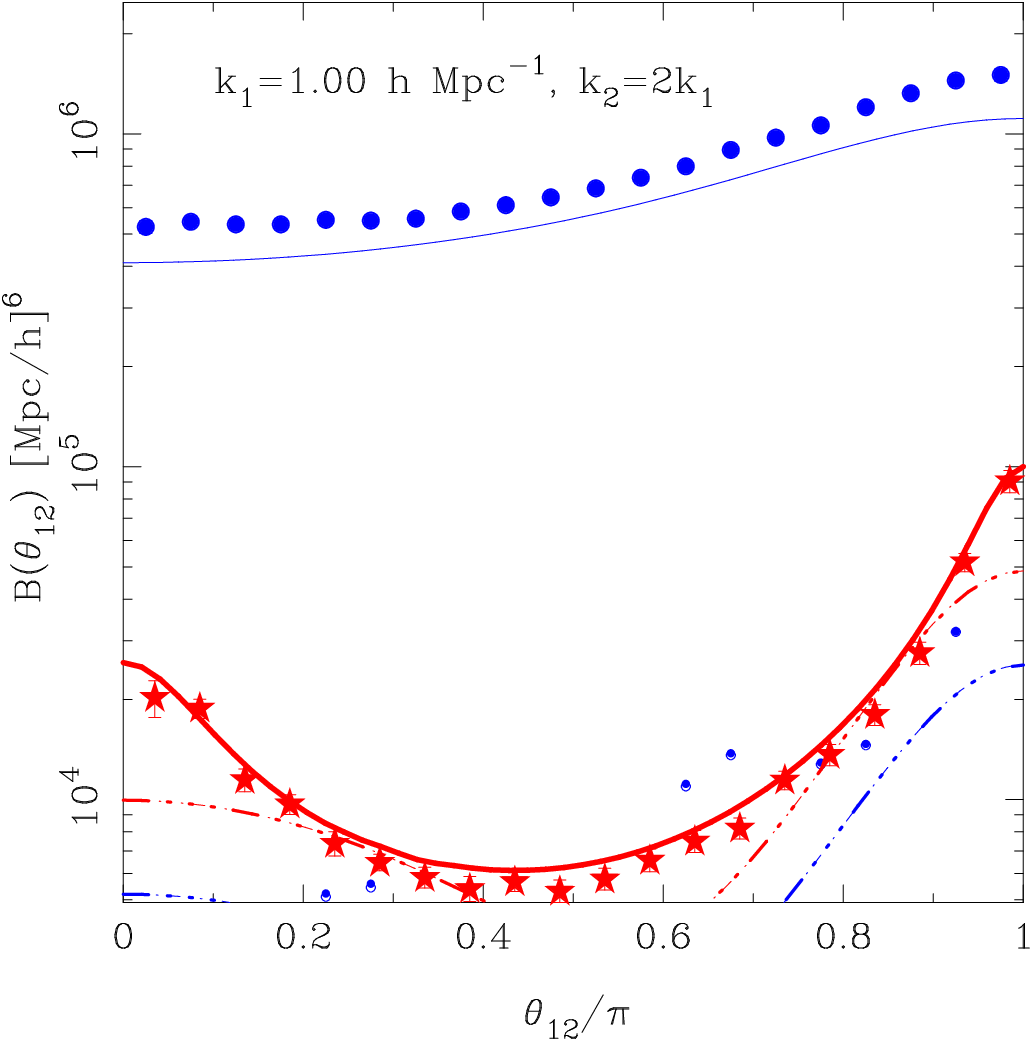}}
\caption{\small{Configuration dependence of the bispectrum in real and
redshift space measured for the dark matter particles in the ensemble
of LCDM simulations. The four panels show the configuration dependence
for $k$-space triangles with $k_2/k_1=2$ and for scales
$k_1=\left\{0.05,0.1,0.5,1.0\right\}h\Mpc^{-1}$. Red and blue colors
distinguish between real and redshift space quantities. The solid
points with error bars show measurements: large solid points are for
the real (dots) and redshift space (stars) monopole bispectra; the
corresponding smaller points are the imaginary bispectra (which should
be zero). The solid lines show the predictions from HM (thin) and HMs
(thick). The triple dot-dash lines show the predictions from PT (thin)
and PTs (thick). In the top left panel, for clarity we have suppressed
the errors on the imaginary bispectra.}
\label{fig:QthetaRed2}}
\end{figure*}


\begin{figure*}
\centerline{
\includegraphics[width=8.cm]{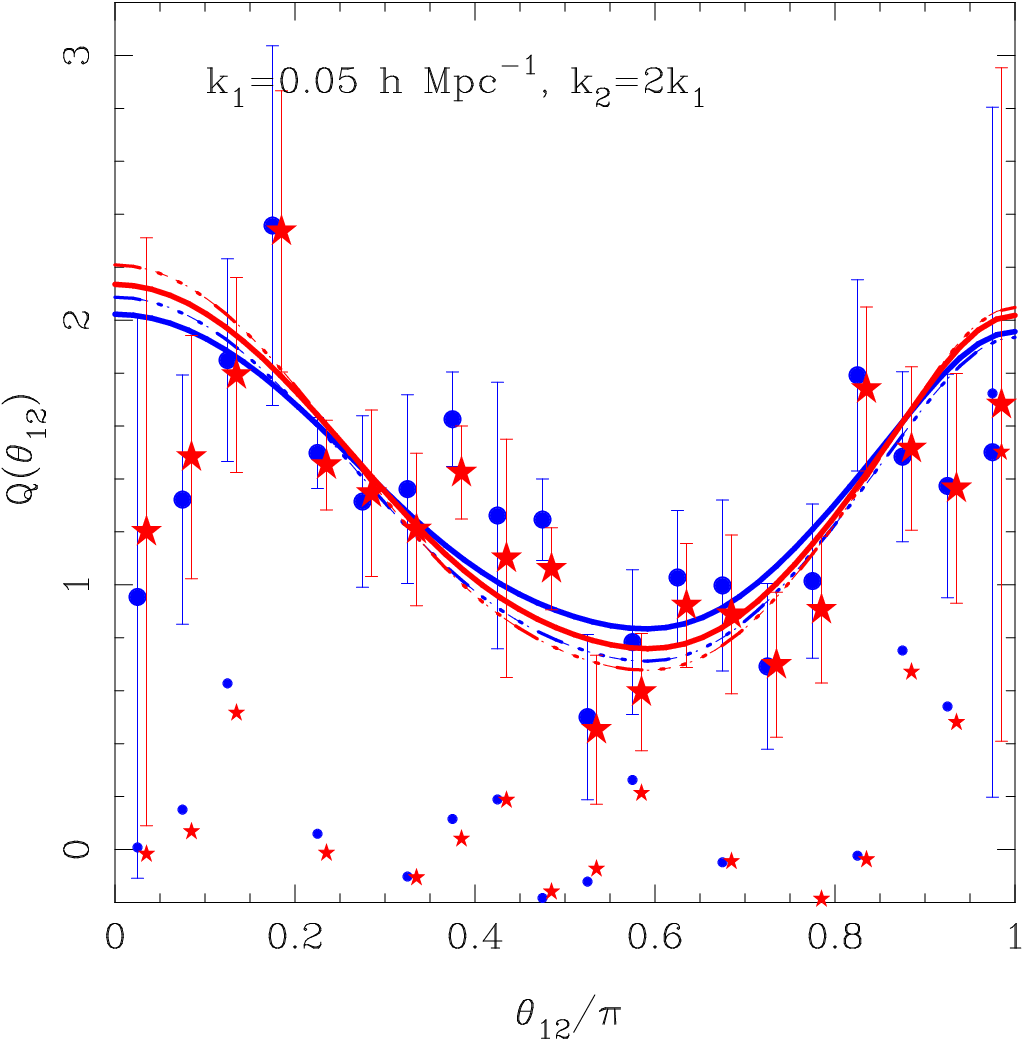}
\includegraphics[width=8.cm]{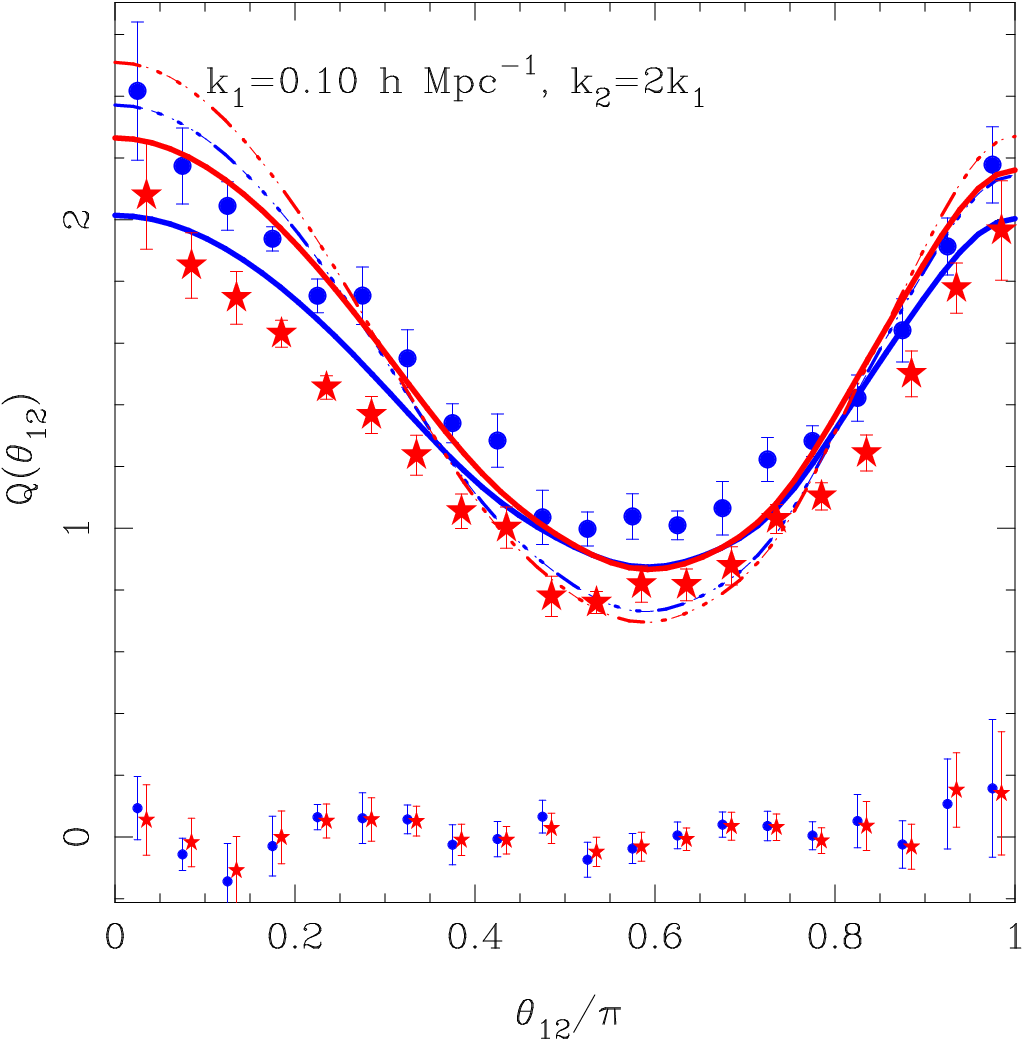}}
\centerline{
\includegraphics[width=8.cm]{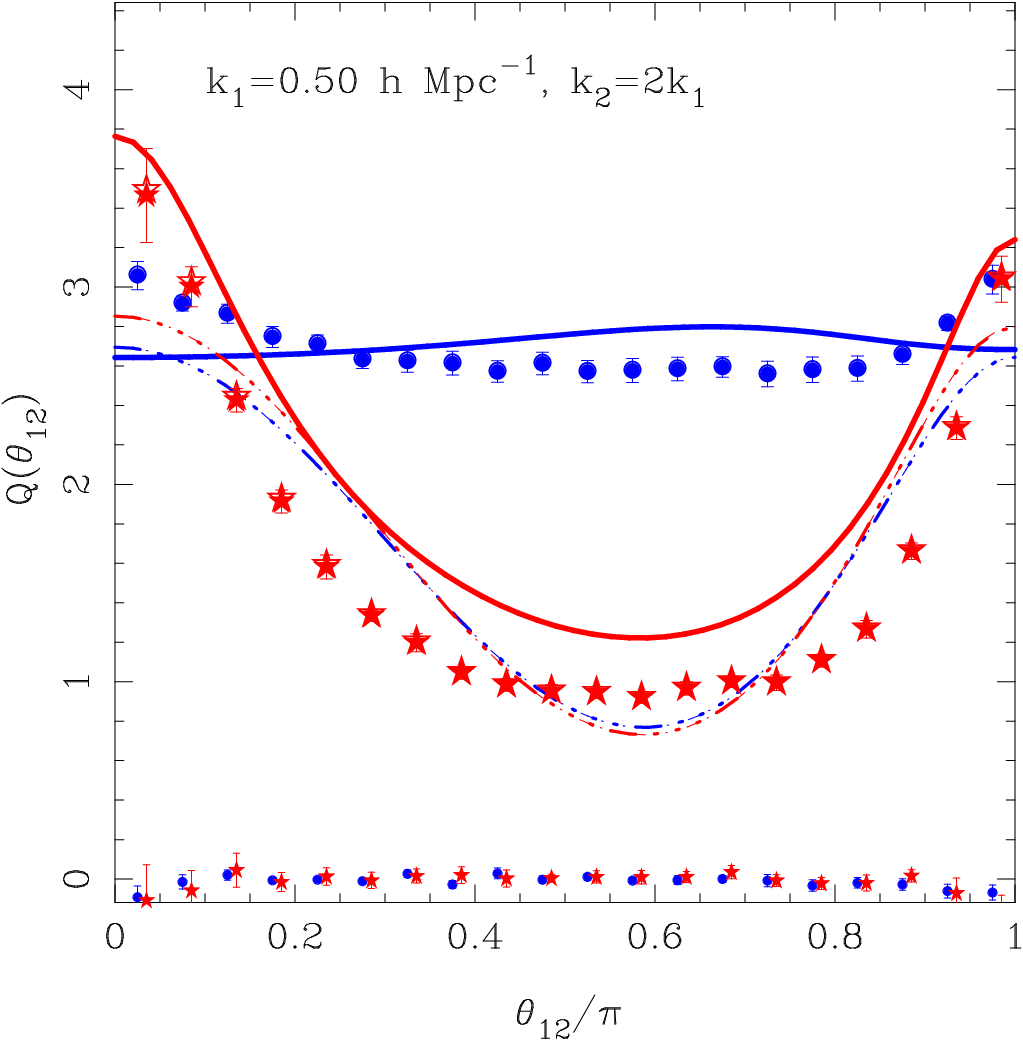}
\includegraphics[width=8.cm]{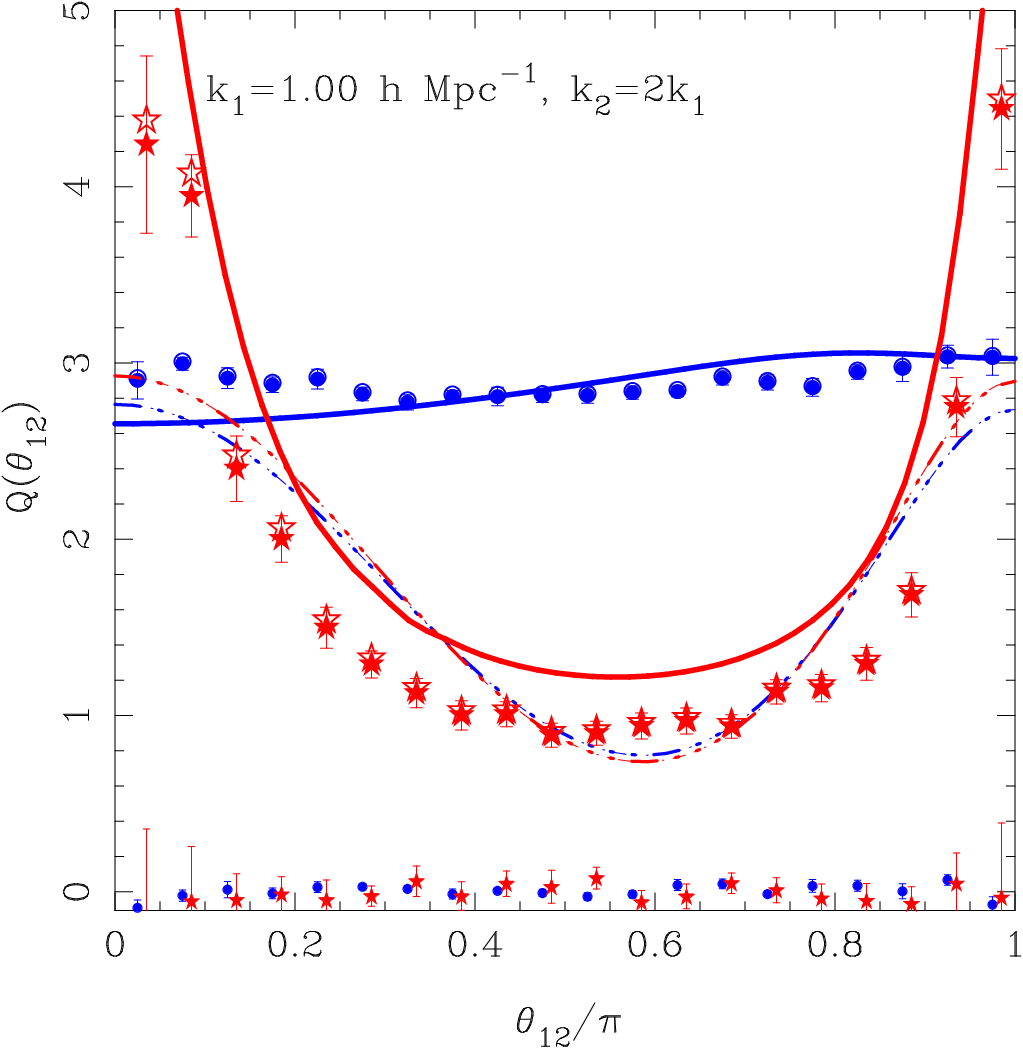}}
\caption{\small{Same as Fig.~\ref{fig:QthetaRed2} only this time for
the reduced bispectrum. Additionally, where visible the open symbols
represent the shot noise corrected estimates.}
\label{fig:QthetaRed3}}
\end{figure*}


\subsection{Bispectrum results}\label{ssec:B}

Figure \ref{fig:QthetaRed2} shows the mean and the 1-$\sigma$ errors
on the mean (i.e. we divide the standard error bars by $1/\sqrt{7}$)
for the configuration dependence of the bispectrum
$B(k_1,k_2,\theta_{12})$ in both real and redshift space for the 8
LCDM simulations. The four panels show results for $k$-space triangles
with $k_2/k_1=2$ and $k_1=\{0.05,0.1,0.5,1.0\} h\Mpc^{-1}$ and thin
(blue) and thick (red) lines distinguish between real and redshift
space quantities.

Considering the largest scales ($k=0.05\, h\Mpc^{-1}$), we see that
the ensemble estimate is rather noisy, this owes to the large sample
variance on these scales. Nevertheless, it can be discerned that the
redshift space (large solid stars) estimate has a slightly higher
amplitude than the real space (solid points) estimate. We also note
that there are a few bins that give a significant non-zero
contribution to the imaginary bispectra. We attribute this to the large
sample variance errors, but in the main these points are below the
true signal.  In all cases the results for the bispectra appear to be
consistent with the PT and HM predictions, to within the errors.  To
say anything more definite on these scales will require much larger
simulation volumes, and we shall defer this question for future study.

On intermediate scales ($k=0.1\, h\Mpc^{-1}$), the estimates for the
real and redshift space spectra are much more significant, the errors
being well defined. This is supported by the fact that the imaginary
spectra have amplitudes that are now too small to be plotted. It can
be seen that the results have a strong dependence as a function of
triangle configuration.  On comparing with the PT
and PTs, we see that on these scales PT under-predicts the measured
quantity by roughly $\sim20\%$, whereas PTs overpredicts the amplitude
by a similar amount. Considering the predictions of the HM and HMs,
it is clear that in real space the model is a significant improvement
over PT; whereas in redshift space, whilst the configuration
dependence appears to have been slightly improved, the amplitude is
still too large.

On smaller scales ($k=0.5\, h\Mpc^{-1}$), we see that the estimates of
the spectra are of even higher signal-to-noise and that they form
tight-loci across the configuration. Again the imaginary bispectra
are insignificant. It may be noticed that there is a large difference
between the real space PT and the simulation estimate -- almost two
orders of magnitude, this is due to the fact that $B \sim Q P^2$ and the power spectrum is significantly larger than linear. However, this dramatic change in the measured
bispectrum is captured relatively well by the HM, which underpredicts
the result by only $\sim20\%$. Turning to the redshift space results,
we are surprised to see that the PTs is of the same amplitude and has
somewhat similar configuration dependence as the estimate. This
agreement seems coincidental, given the poor agreement in real
space. In reality the true dynamics on these scales must already be
very non-linear. Lastly, we draw attention to the fact that our HMs
result is in excellent agreement with the simulation data.

Examining scales on the order of the virial radii for clusters
($k=1.0\, h\Mpc^{-1}$), we see again that the estimates are of very
high significance. Again, the HMs predictions are in excellent
agreement.  This essentially vindicates our form for the 1-Halo term,
and means that the configuration dependence is very much governed by
the FOG distortions.


\subsection{Reduced bispectrum results} 

Figure \ref{fig:QthetaRed3} shows similar results as in
Fig.~\ref{fig:QthetaRed2} but for the configuration dependence of the
reduced bispectrum, $Q(k_1,k_2,\theta_{12})$. Again errors are shown
on the mean of $Q$.

On the largest scales probed, ($k=0.05\, h\Mpc^{-1}$), the estimates
are noisy, but there is evidence for an excess of signal for co-linear triangles --
meaning that on average structures are more filamentary than spherical
on the largest scales \cite{Scoccimarroetal1998,Scoccimarroetal1999}.
Also, the data appear to be scattered about the theoretical
predictions, with all models being equally good fits to the data. 

Considering intermediate scales ($k=0.1\, h\Mpc^{-1}$), the estimates
are much more significant and possess well defined configuration
dependencies -- both showing an excess of signal for colinear
triangles. However, the real space bispectrum appears to be in excess
of the redshift space quantity. This is in contrast to the PT and PTs
predictions which, whilst qualitatively capture the overall shape,
predict the reverse trend, the results being discrepant by roughly
$\sim20\%$. This problem is also mirrored in the HM and HMs
predictions, but the flatter configuration dependence of the data is
better captured by the Halo Model. As was discussed in
Sec.~\ref{ssec:largescalelimit}, this flattening can be attributed to
the impact of the 1- and 2-Halo terms acting as effective shot noise
contributions. The amplitude offsets still require explanation, and we
refer to our discussion of Sec.~\ref{ssec:estpower} for some possible
remedies, but to that list we may now add the need for loop
corrections to the tree-level bispectra. As was shown in
\cite{Scoccimarroetal1998}, in real space the 1-Loop corrections are
significant on these large scales. 

On smaller scales ($k=0.5\, h\Mpc^{-1}$), we see that, as was noted in
Fig.~\ref{fig:QthetaRed3}, the real space measurements have increased
in amplitude and have become much flatter across the configuration --
the HM predictions agree rather well with this result. This implies
that the statistic is already in the fully non-linear regime, since
the 1- and 2-Halo terms are dominating the signal here. There is very
little relation between standard PT to the measurements, as expected.  Turning to
the redshift space estimates, it can be seen that the configuration
dependence displays a reasonably strong $U$-shape. The PTs predictions
do not describe this shape very well, but are not as discrepant as in
real space. However, HMs predictions capture the form of the
configuration dependence exceptionally well, but are offset by $\sim
10-20\%$.

Considering the scales associated with the virial radii of clusters
($k=1.0\, h\Mpc^{-1}$), we see that the estimates in real space are
surprisingly unchanged and that the HM still provides a very good
description of the data. For the redshift space estimates, we find
that there is now a very strong $U$-shape configuration dependence, in
full agreement with the results from
\cite{Scoccimarroetal1999}. Again, the HMs predictions capture this
result remarkably well, although there is a small amplitude offset.
It is believed that this may be mitigated by implementing the
improvements discussed in Sec.~\ref{ssec:B} and
Sec.~\ref{ssec:estpower}.

We also note that the small discrepancies between the real space HM
predictions and the simulations, are entirely consistent with the work
of \cite{SmithWattsSheth2006} -- in reality haloes are triaxial rather
than spherical, and this shows up as a characteristic increase in
signal for colinear triangles and a suppression for isosceles
configurations.


\subsection{Code comparison}

Owing to the algorithmic differences between the bispectrum estimation
procedure presented in Section \ref{ssec:estimation} and that
presented in \cite{Scoccimarro2000}, we decided to compare the results
obtained from these two approaches. Besides providing an important
cross-check, this also enables us to examine how accurately the two
codes are recovering $Q$. Overall we found very good agreement between
both methods, and full results are presented in Appendix
\ref{app:codecomp}.


\section{Conclusions}\label{sec:conclusions}

In this paper, we have provided a new analytic model for the redshift
space bispectrum of dark matter, haloes and galaxies in the plane
parallel approximation for the redshift space distortion.  On large
scales, the model predictions have a direct correspondence to the
non-linear perturbation theory and on small scales, the predictions
are entirely governed by the phase space density of galaxies/dark
matter internal to the haloes. This is the first time that the
information from bulk flows and virial motions have been naturally
incorporated into an analytic model for the higher-order clustering
statistics in redshift space.

In our analytic model, the bispectrum is represented as a sum over
three terms; these correspond to all of the possible distinct
arrangements of three points in three haloes, and we referred to these
as the 1-, 2- and 3-Halo terms. A practical evaluation of the
monopole of the bispectrum (a direct observable), with realistic models
for halo profiles, abundance and clustering, required the execution of
a set of 4-D numerical integrals. For the terms that involved
large-scale correlations (2- and 3-Halo terms) it was shown that these
expressions could be easily modularized, and so are best computed in
parallel. The 1-Halo term must be integrated with an efficient
higher-dimensional integrator.

It was shown that the large-scale predictions in the model, which are
governed by the 2- and 3-Halo terms, can be simplified greatly under
the approximation that the angle average of the product of anisotropic
density profiles and large scale clustering of halo centers can be
performed separately \cite{White2001,Seljak2001}.  This approximation
was accurate to $<10\%$ on scales $k=0.1 h/\Mpc$, elsewhere it was of
the order $<5\%$.

The predictions for the bispectrum monopole were compared with
analytic PT in real and redshift space. On very large scales
$k=0.05-0.1 h\Mpc^{-1}$, the model closely agreed with the redshift
space PT predictions, but with some small deviations noticeable. It
was argued that these were due to the `effective' shot-noise like
behavior of the 1- and 2-Halo terms in the low-$k$ limit. On smaller
scales $k=0.5-1.0 h\Mpc^{-1}$ the analytic model showed a dramatic
departure from the PT predictions and displayed a {\em U}-shaped
anisotropy \cite{Scoccimarroetal1999,GaztanagaScoccimarro2005}. This
was the imprint in the configuration dependence of the FOG distortions
from non-linear virial motions. No trace of the PT remained in the
model predictions on these scales. The model predictions showed that
there was no scale where a hierarchical model provided a good
description of the configuration dependence of the bispectrum.

The predictions were then confronted with measurements of the
bispectrum and reduced bispectrum from an ensemble of numerical
simulations. On very large scales $k=0.05\,h\Mpc^{-1}$ it was found
that the PT and Halo Model predictions were equally good, to within
the errors. On smaller scales, $k=0.1\,h\Mpc^{-1}$, departures
between PT and the simulations were noted at the level of
$\sim10-20\%$. Therefore, studies that use the lowest order PT
to extract galaxy bias are unlikely to be robust on scales $k \ga 
0.1\,h/\Mpc$. The Halo Model was a better description of the data, but
was not in perfect agreement. Plausible improvements to the Halo
Model on these scales were discussed. On even smaller scales,
$k=0.5-1.0\,h\Mpc^{-1}$, the configuration dependence was flat in real
space, whereas in redshift space there was a very strong $U$-shape
feature. The numerical results were reasonably well reproduced by
our halo model predictions, a significant improvement over PT that 
breaks down at these scales. 

We cross-validated our results from the numerical simulations by
comparing our bispectrum estimates with those obtained from the
independent code of R.~Scoccimarro. The results from the two codes
were found to be in very good agreement, although our results were a
factor of 2-3 times more noisy on the largest of scales.

In future work, we shall extend our analysis to examine the halo and
galaxy clustering as a function of mass and type. It will be also
important to resolve whether or not the Halo Model's effective
shot-noise terms are important for modeling real survey data. Owing to
the fact that different galaxy populations are easily and naturally
included into our description, it is hoped that this approach will
help to facilitate extraction of information from current and future
hi-fidelity large scale structure surveys of the Universe.


\section*{Acknowledgments}

The authors thank Uros Seljak, Bhuvnesh Jain, Laura Marian, Cameron
McBride and Bob Nichol for useful discussions. We thank Rob Thacker
for advice and encouragement regarding the opteron cluster. We thank
Volker Springel for making public his {\tt GADGET-2} code. RES
acknowledges support from the Swiss National Foundation.  RES and RKS
acknowledge support from the National Science Foundation under Grant
No. 0520647.  RS is partially supported by NSF AST-0607747 and NASA
NNG06GH21G.


\begin{widetext}

\appendix


\section{Duality of clustering statistics}\label{sec:theory}

This section defines the configuration and Fourier space clustering
statistics and their dual relationship with one another.  It also
shows how the cosmological assumptions impose certain important
conditions upon these statistics.

We shall assume that ensemble averages are taken over a volume, $\Vu$,
of the Universe sufficiently large for the fundamental $k$-space cell
volume to be considered infinitesimally small; outside of this large
volume our stochastic fields are exactly zero.  This allows us to
define Fourier transforms in the volume.  We shall also let these
fields be Ergodic, whence ensemble averages are equivalent to averages
over volume.


\subsection{Real space representation: correlation functions}

The fractional density field of matter is defined:
\be \delta(\bx,t)\equiv [\rho(\bx,t)-\rhob(t)]/\rhob(t) \ ,\ee
where $\rho(\bx,t)$ is the local density at coordinates $(\bx,t)$ and
$\rhob(t)$ is the density of the homogeneous background at time $t$.
The $n$-point connected auto-correlation functions of $\delta$,
represent the excess probability from the product of the individual
independent 1-point distributions of obtaining a particular set of
values at all points, e.g. the probability of obtaining fluctuations
at three points $(a,b,c)$ can be written:
\be P(a,b,c)\equiv
P(a)P(b)P(c)\left[1+C_2(a,b)+C_2(b,c)+C_2(c,a)+C_3(a,b,c)\right] \
\label{eq:jointpdf} .\ee
where $C_2$ and $C_3$ are the connected 2-- and 3--point correlation
functions. We may now be clear about what we mean by connected
correlation function: the connected correlator may not be reduced to
sums over products of lower order connected correlators. In cosmology
these are more commonly written:
\ba 
\xi_2(\bx_1,\bx_2|t) & \equiv &
\left<\delta(\bx_1,t)\,\delta(\bx_2,t)\right>_c\ ;\label{eq:xi2}\\
\xi_3(\bx_1,\bx_2,\bx_3|t) & \equiv &  \left<
\delta(\bx_1,t)\,\dots\,\delta(\bx_3,t)\right>_c\
\label{xi3deff} \ ; \label{eq:xi3}\\
\xi_n(\bx_1,\dots,\bx_n|t) & \equiv & \left<
\delta(\bx_1,t)\,\dots\,\delta(\bx_n,t)\right>_c\ \label{xindeff}
\label{eq:xin} \ .
\ea

These functions obey an integral constraint
\be \frac{1}{\Vu}\int \dx_n \xi_n(\bx_1,\dots,\bx_n) \rightarrow 0 \ ;\ n>1\ .\ee
This follows from noting that on marginalizing the probability
functions over one variable, say the $N$th variable, one finds that
the resulting distribution depends on $n-1$-points, and therefore
from Eq.~(\ref{eq:jointpdf}) must not depend on the $n$-point
connected correlation function.

If the density field obeys the cosmological principle, that is
statistical homogeneity and isotropy on scales greater than the
coherence scale of our fields, then the correlation functions are
invariant under translation and rotation of the coordinate
system. They are also parity invariant real functions and are
invariant to exchange of vector arguments. Thus:
\ba 
\xi_n(\bx_{1},\dots,\bx_{n})  
& = & 
    \xi_n(\bx_{1}+\bx_0,\dots,\bx_{n}+\bx_0)  
    \hspace{0.5cm} {\rm (Translation)} \ ;\label{eq:xiTrans}\\
& = & 
    \xi_n({\mathcal R}\bx_{1},\dots,{\mathcal R}\bx_{n}) 
    \hspace{1.5cm} {\rm(Rotation)}\ ;\label{eq:xiRot}\\ 
& = & 
    \xi_n(-\bx_{1},\dots,-\bx_{n}) 
    \hspace{1.6cm} {\rm (Parity)}\ ; \label{eq:xiPar}\\
& = & 
    \xi_n(\bx_{2},\bx_{1},\dots,\bx_{n}) 
    \hspace{1.6cm} {\rm (Exchange)}\ ;\label{eq:xiExch} \nonumber \\
& = & \xi_n(\bx_{i},\bx_{2},\dots\bx_1,\dots,\bx_{n}) 
\ea
For anisotropic fields rotation invariance is broken and for
inhomogeneous fields translation symmetry is broken. For homogeneous
fields we may immediately apply the translational invariance and drop
one of the vector arguments in our function. Setting ${\bx_0=-\bx_n}$
in Eq.~(\ref{eq:xiTrans}) gives,
\be \xi_n(\bx_{1},\dots,\bx_{n})= \xi_n(\bx_{1n},\dots,\bx_{(n-1)n}) \
    ,\ee
where $\bx_{ij}\equiv\bx_i-\bx_j$ and we shall not write the zero
argument in the $n$th space. In this paper we will mainly be concerned
with clustering statistics that obey homogeneity, but are anisotropic,
as this is exactly the case for the redshift space distortion in the
plane parallel approximation.  Lastly, we have the closure relation:
$\bx_{21}+\bx_{32}+\dots+\bx_{1n}={\bf 0}$.


\subsection{Fourier space representation: poly-spectra} 

Under the conditions stated earlier, the density field $\delta(\bx,t)$
may be equivalently written as an infinite sum over plane waves through
the Fourier transform, where our Fourier convention is
\be \delta(\bx)=\frac{\Vu}{(2\pi)^3}\int \dk\delta(\bk)e^{-i\bk\cdot\bx}
\ \Leftrightarrow\ \ \delta(\bk)=\frac{1}{\Vu}\int
\dx\,\delta(\bx)e^{i\bk\cdot\bx}\ .\ee
Transforming the density terms in Eqs~(\ref{eq:xi2}--\ref{eq:xin}),
leads to
\ba \xi_2(\br_{12}) & = & \int \prod_{i=1}^{1}
\left\{\frac{\dk_i}{(2\pi)^3}\right\} P_2(\bk_1,\bk_2) 
e^{-i\bk_1\cdot \br_{12}}
\ ; \hspace{0.5cm}\left[\bk_1+\bk_2={\bf 0}\right]\\
\xi_3(\br_{13},\br_{23}) & = & \int \prod_{i=1}^{2}
\left\{\frac{\dk_i}{(2\pi)^3}\right\} 
P_3(\bk_1,\bk_2,\bk_3) e^{-i\bk_1\cdot \br_{13}-i\bk_2\cdot\br_{23}} \
; \hspace{0.5cm} \left[\bk_1+\bk_2+\bk_3={\bf 0}\right] \label{eq:xi3Bispec}\\
\xi_n(\br_{1n},\dots,\br_{(n-1)n}) & = & \int \prod_{i=1}^{n-1}
\left\{\frac{\dk_i}{(2\pi)^{3}}\right\} 
 P_n(\bk_1,\dots,\bk_n)
e^{-i\bk_1\cdot \br_{1n}-\dots-i\bk_{(n-1)}\cdot\br_{(n-1)n}} 
\ ;\hspace{0.5cm}\left[\sum_{i=1}^{n}\bk_i={\bf 0}\right]\ea
where we have very generally defined the $n$-point spectrum as:
\be \Vu^{n-1}\left<\delta^s(\bk_1)\dots\delta^s(\bk_n)\right> = 
P_n^s(\bk_1,\dots,\bk_n)\left[\delta^D_{1\dots n}\right](2\pi)^3/\Vu\ ,
\ee
this condition simply arises from the imposed harmonic boundary
conditions within our volume: only overlapping waves are constructive.
The short hand notation
$\left[\delta^D_{12\dots}\right]\equiv\delta^D(\bk_1+\bk_2+\dots)$ has
been adopted for the argument of the Dirac delta function. The
presence of this term ensures that the sum of $k$-vectors forms a null
vector, $\bk_1+\dots+\bk_n={\bf 0}$ and we shall refer to this as the
closure condition. For the case of $(n=2)$, we have the power spectrum
$P_2(\bk_1,\bk_2)\equiv P(\bk_1)$ and for $(n=3)$ we have the
bispectrum $P_3(\bk_1,\bk_2,\bk_3)\equiv B(\bk_1,\bk_2,\bk_3)$.

Conversely, the power spectra may also be written as inverse Fourier
transforms of the correlation functions:
\ba 
P(\bk_1) & = & \int d^3\!\br_{12}
\,\xi(\br_{12}) e^{i\bk_1\cdot\br_{12}}\
\label{eq:Pow2Cf2}; \\
B(\bk_1,\bk_2,\bk_3) & = & 
\!\!\int d^3\!\br_{13}\, d^3\!\br_{23}\, \zeta(\br_{13},\br_{23})
\,e^{i\left[\bk_1\cdot{\br_{13}}+\bk_2\cdot\br_{23}\right]}
\label{eq:Bi2Cf3} ,\\
P_n(\bk_1,\dots,\bk_n) & = & 
\!\!\int d^3\!\br_{1n}\,\dots d^3\!\br_{(n-1)n}\, 
\xi_n(\br_{1n},\dots,\br_{(n-1)n})\,e^{i\left[\bk_1\cdot{\br_{1n}}+\dots+\bk_{(n-1)}\cdot\br_{(n-1)n}\right]}
\label{eq:Polyspectra} \ .
\ea
From these relations and the properties of the correlation functions,
Eqs~(\ref{eq:xiTrans}--\ref{eq:xiExch}), we may now infer the
corresponding properties for the poly-spectra. Translational
invariance in configuration space means that the poly-spectra are
invariant under a phase shift to the Fourier density fields.
Invariance to rotation of the coordinate frame leads to rotation
invariance of the polyspectra:
\ba 
P_n(\bk_1,\dots,\bk_n) & = & \int d^3\![{\mathcal R}\br_{1n}]\,\dots
d^3[{\mathcal R}\!\br_{(n-1)n}]\,
\xi_n({\mathcal R}\br_{1n},\dots,{\mathcal R}\br_{(n-1)n})
\,e^{i\left[\bk_1^T{{\mathcal R}\br_{1n}}+\dots+
\bk_{(n-1)}^{T}{\mathcal R}\br_{(n-1)n}\right]} \ ,\nonumber \\
 & = & \int d^3\!\br_{1n}\,\dots
d^3\!\br_{(n-1)n}\,
\xi_n(\br_{1n},\dots,\br_{(n-1)n})
\,e^{i\left[\br_{1n}^{\rm T}{\mathcal R}^{\rm T}\bk_1+\dots+
\br_{(n-1)n}^{\rm T}{\mathcal R}^{\rm T}\bk_{(n-1)}\right]} \nonumber \\
 & = & P_{n}(\mathcal R^{\rm T}\bk_1,\dots,\mathcal R^{\rm T}\bk_n)
\ea
Parity invariance and the reality of the configuration space functions
leads to the reality of the poly-spectra:
\be P_n(\bk_1,\dots,\bk_n)  =  P_{n}(-\bk_1,\dots,-\bk_n) = 
\left[P_n(\bk_1,\dots,\bk_n)\right]^{*}\ , \label{eq:realspectra}\ee
where the $*$ corresponds to complex conjugation.  Many of these
properties simplify the analysis in the main text.


\section{Calculational details}\label{sec:details}


\subsection{The NFW density profile with the Bullock et al. normalization}
\label{ssec:profiles}

As described in Section \ref{sec:RedDen} to compute the redshift space
density profile we require a model for the real space density profile
$\rho(r)$ and a model for the 1-point velocity distribution function
of particles in a halo. For the density profile we adopt the `NFW'
model \cite{NavarroFrenkWhite1997}:
\be \rho(r)=\rho_c\left[y(1+y)^2\right]^{-1}\ ; \ \ \ y\equiv
r/r_c \ .\ee
This model is fully determined by two parameters, $\rho_c$ and $r_c$,
a characteristic density and radius. These two parameters are not
independent, but are related by the mass enclosed:
\be \rho_c = \frac{\rhob \Delta_{\vir}c^3/3}{\log(1+c)-c/(1+c)}
\ ;\ \ c\equiv \frac{r_{\vir}}{r_c}\ ,\ee
where $c$ is the concentration parameter and is the ratio of the
virial radius to the characteristic radius. The virial radius is the
boundary layer within which all particles have undergone violent
non-linear relaxation. It is taken to be specified through
\be M_{\vir}=\frac{4}{3}\pi r_{\vir}^3\Delta_{\vir}\rhob\ ,
\label{eq:Mvir}\ .
\ee
$\Delta_{\vir}$ is the density contrast for virialization, which may
be estimated from the spherical collapse model. For flat universes
with a cosmological constant a good fit to the functional form is
provided by \cite{BryanNorman1998}
\be \Delta_{\vir}=\left[(18\pi^2+82x-39x^2\right]/\Omega(a)\ ;\ \ \ 
x\equiv[\Omega(a)-1]\ .\ee

To obtain the concentration parameter as a function of mass we follow
the model of Bullock et al. \cite{Bullocketal2001}. For this we have
\be c=K\frac{a}{a_{c}(M_{\vir})}\ee 
where we take $K=3.0$ and where $a_c(M_{\vir})$ is the collapse
expansion factor for a halo of mass $M_{\vir}$. The collapse expansion
factor for a halo of mass $M$ may be determined through solving the
relation
\be \frac{D_1(a_c)}{D_1(a_0)}\sigma(FM_{\vir},a_0)=\delta_c \ ,\ee
where $F=0.001$ is taken as a fixed fraction of the initial mass,
$D(a)$ is the linear theory growth factor at epoch $a$. The parameter
$\delta_c=1.686$ is the linearly extrapolated density threshold for
collapse from the spherical collapse model, where we ignore the slight
dependence on cosmology \cite{Lahavetal1991}. As was shown by
\cite{Bullocketal2001} this model provides a very good description of
the ensemble average properties of dark matter haloes. More
sophisticated models may be constructed that take into account that
haloes are more complicated, e.g. a halo of mass $M$ drawn at random
from the ensemble will have a concentration parameter that is drawn
from a probability distribution of possible concentrations. In
addition one may include sub-structure \cite{ShethJain2003} or halo
triaxiality \cite{SmithWatts2005,SmithWattsSheth2006}. However, we
shall leave these additional embellishments for future study.


\subsection{Conversion between Sheth \& Tormen and Bullock et al mass defintions}\label{ssec:MassConversion}

The definitions of halo mass that are used in the Sheth \& Tormen mass
function and the Bullock et al. model for the density profile are
inconsistent. We recall that the Sheth \& Tormen halo mass is defined:
\be M_{\rm ST}=\frac{4}{3}\pi r_{\rm ST}^3 200\rhob \ .\ee
We resolve this inconsistency using the methodology of
\cite{SmithWatts2005}, which briefly is as follows: For a given halo
of the NFW type, the physical values of the characteristic density and
radius are independent of our specific choice of halo mass. Using the
relation for the physical density as a constant we arrive at the
mapping
\be \left(\frac{c_{\vir}}{c_{\ST}}\right)^3=\frac{200}{\Delta_{\vir}}
\left[\frac
{\log(1+c_{\vir})-c_{\vir}/(1+c_{\vir})}
{\log(1+c_{\ST})-c_{\ST}/(1+c_{\ST})}\right]
\ .
\ee
Thus if we take a Bullock et al. mass and derive the appropriate
$c_{\vir}$ we may then solve the above expression to find
$c_{\ST}$. Following this we may then obtain the corresponding Sheth
\& Tormen mass through application of the relation,
\be M_{\ST}=\frac{200}{\Delta_{\vir}}\left(\frac{c_{\ST}}{c_{\vir}}\right)^3
M_{\vir}\ .\ee
%


\subsection{1-Point velocity distribution profile}

For the 1D velocity distribution function we adopt the standard
Maxwellian distribution \cite{Sheth1996,White2001,Seljak2001,Kangetal2002}:
\be
{\mathcal V}[u_z|\sigma_{\D1}(M_{\vir})]du_z=\frac{1}{\sqrt{2\pi}\sigma_{\D1}}
\exp\left[-\frac{u_z^2}{2\sigma_{\D1}^2}\right]du_z \ ,\ee
where $\sigma_{\D1}$ is the 1D velocity dispersion. For haloes that
possess an isothermal density distribution, this quantity is related
to the halo circular velocity ($V_c$) through the following relation
\cite{BinneyTremaine1988}
\be \sigma_{\D1}^2(M_{\vir})= \epsilon V_c^2/2\
; \hspace{0.5cm} 
V_c^2=\frac{GM_{\vir}}{r_{\vir}}\label{eq:isodisp}\ee
Note that we have included a parameter $\epsilon$ into
Eq.~(\ref{eq:isodisp}), this may be used to account for the fact that
the relation is only approximately true for the NFW density profile
model. It also serves the further purpose of allowing us to turn off
the fingers-of-god through setting $\epsilon\rightarrow \eta\ll 1$.
As discussed in Section \ref{ssec:estpower} and shown in Fig
\ref{fig:SigmaMass} $\epsilon=0.76$ provides a reasonable fit to the
velocity dispersion mass relation from simulations. On combining the
above relations we have
\be \sigma_{\D1}^2(M_{\vir}) = \frac{100\,\epsilon}{2}\, \Omega_m(a)\,
 \Bigl[H(a)\,r_{\vir}\Bigr]^2 
\label{eq:sigma}\ , \ee
Notice that the ratio $\sigma_{\D1}(M_{\vir})/r_{\vir}$ is independent
of halo mass. One immediate consequence of this is that in redshift
space the ratio of the line-of-sight projection of particles in a halo
compared to the transverse length will be a constant,
$\sigma_{\D1}/(Hr_{\rm vir})\sim3$, regardless of mass.  In other
words, FOGs lead to density profiles which are self-similar. In our
analysis we take $\epsilon=0.76$ (See Fig.~\ref{fig:SigmaMass}).
Finally, the Fourier transform of the velocity distribution is
\be {\mathcal V}(\mu_1k_1)=\exp\left\{-\frac{1}{2}[k_1 \mu_1
\sigma_{\D1}(M_{\vir})]^2\right\}\ .\label{eq:1ptveldist}\ee


\begin{figure*}
\centerline{ \includegraphics[width=8cm]{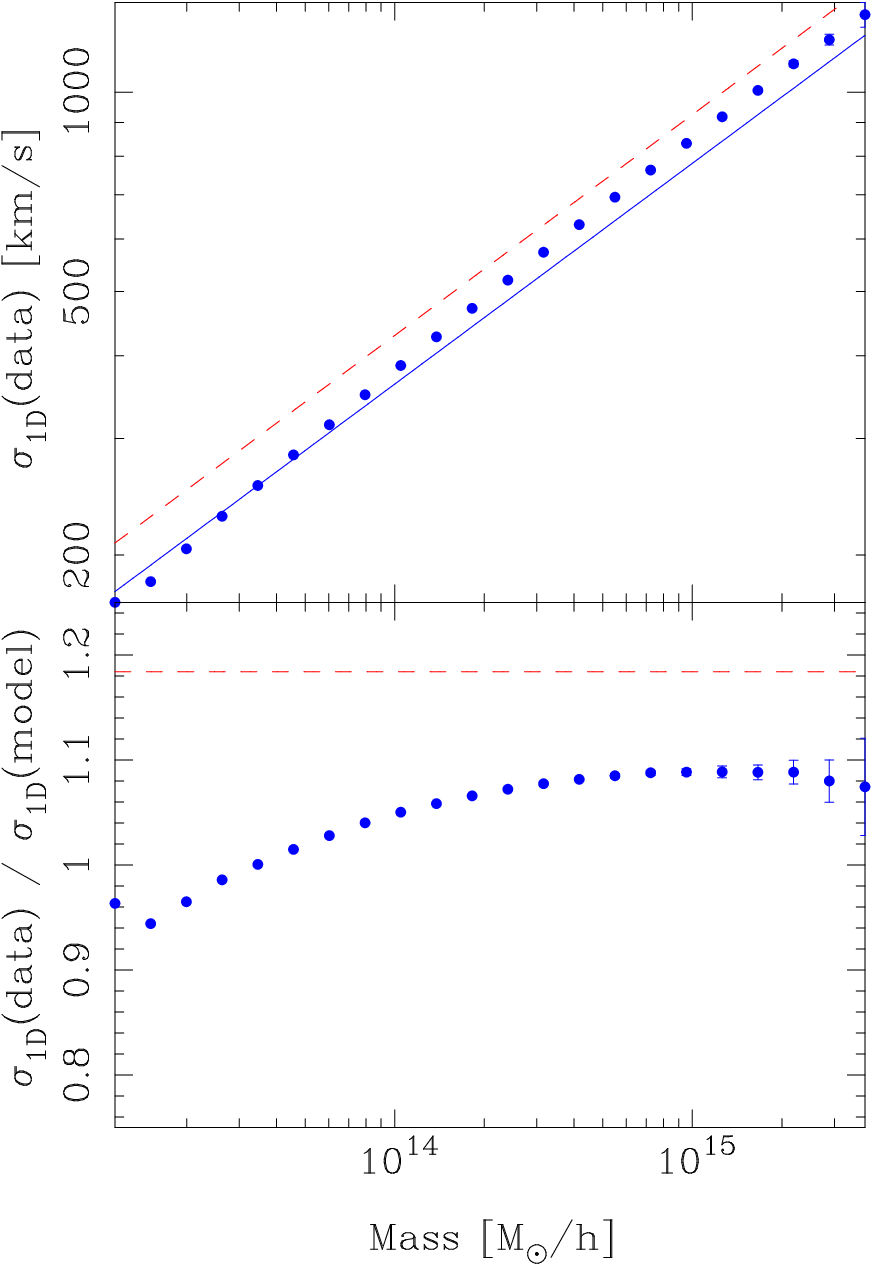}}
\caption{\small{Mass verses 1D velocity dispersion measured for FoF
haloes in simulations. Points show mean and 1-sigma errors for
measurements from the numerical simulations. The solid and dash line
shows the predictions from Eq.~\ref{eq:sigma} with
$\epsilon=\{0.76,1.0\}$. The bottom panel shows the ratio of the data
with respect to the $\epsilon=0.76$ model.}
\label{fig:SigmaMass}}
\end{figure*}


\section{Eulerian PT and Halo-PT kernels}\label{sec:HaloPTKernels}

The first two symmetrized Eulerian PT kernels for the density and
divergence of the velocity field are \cite{Bernardeauetal2002}:
\ba 
F_{1,2} & = & \frac{5}{7}+\frac{\mu_{12}}{2} \left[\frac{q_1}{q_2}
+\frac{q_2}{q_1}\right]+\frac{2(\mu_{12})^2}{7} \ ; \\
G_{1,2} & = & \frac{3}{7}+\frac{\mu_{12}}{2}
\left[\frac{q_1}{q_2}+\frac{q_2}{q_1}\right]+\frac{4(\mu_{12})^2}{7} 
\ ,\ea
where $F_{1,\dots,j}\equiv F_j(\bq_1,\dots,\bq_j)$ and where
$\mu_{12}\equiv \bq_1\cdot\bq_2/q_1q_2$.

The Halo-PT kernels, symmetrized in all of their arguments, may be
written in terms of the Eulerian PT kernels up to 2nd order as
\cite{SmithScoccimarroSheth2007}:
\ba 
F^{\hc}_{0}   & = & \frac{}{} b_0(M) \label{eq:HaloPT-K0}\ ;\\ 
F^{\hc}_{1}   & = & \frac{}{} b_1(M)W(|\bk|R) F_{1} \label{eq:HaloPT-K1}\ ; \\ 
F^{\hc}_{1,2} & = & \frac{}{} b_1(M)W(|\bk|R)F_{1,2} + 
\frac{b_2(M)}{2}W(|\bq_1|R)W(|\bq_2|R) F_1 F_2 \label{eq:HaloPT-K2}
\ea
where $F^{\hc}_{1,\dots,j}\equiv F_j^{\hc}(\bq_1,\dots,\bq_j|M,R)$ and
where $\bk=\bq_1+\dots+\bq_j$.
%


\section{Rotation matrix}\label{app:rotation}

Owing to there being several equivalent ways to define the Euler
angles and hence the rotation matrix $\RotII$, we make explicit our
adopted choice. The angles are defined as follows: $\gamma_1$
describes a rotation of the coordinate system around the $z$-axis;
$\gamma_2$ a rotation around the new $y'$-axis; and $\gamma_3$ a
rotation around the new $z''$-axis (\cite{MathewsWalker1970}). Thus,
the components of any vector $\bk$ specified in some initial Cartesian
system can be transformed into the scalar components of the new
rotated basis vectors through $\bk'=\RotII\bk\label{eq:rotk}$.  The
$z-y'-z''$ rotation matrix is \cite{MathewsWalker1970}:
\be \RotII\equiv \left(
\begin{array}{ccc}
\left[ C{\gamma_2}\, C{\gamma_1} C{\gamma_3}-S{\gamma_1}\,S{\gamma_3}\right]
& 
\left[ C{\gamma_2}\, S{\gamma_1}\,C{\gamma_3}+C{\gamma_1}\,S{\gamma_3}\right] 
& 
-S{\gamma_2}\, C{\gamma_3} \\
\left[-C{\gamma_2}\, C{\gamma_1}\, S{\gamma_3} -S{\gamma_1}\,C{\gamma_3}\right]
& 
\left[-C{\gamma_2}\, S{\gamma_1}\,S{\gamma_3}+C{\gamma_1}\,C{\gamma_3}\right]
& 
S{\gamma_2}\, S{\gamma_3} 
\\
S{\gamma_2}\, C{\gamma_1} 
& 
S{\gamma_2}\, S{\gamma_1} 
& 
C{\gamma_2} \\
\end{array}\right)
\ ,\ee
%
%
%
and we employed the economic notation $C{x}=\cos x$ and $S{x}=\sin
x$.


\section{Redshift space power spectrum monopole in the halo model}

The redshift space power spectrum of tracer particles $\alpha$, 
can be written in the linear halo model as  
\ba P_{\alpha}^s(\bk) & = &
P^s_{\alpha,\1H}(\bk)+P^s_{\alpha,\2H}(\bk)\ :\\
P^s_{\alpha,\1H}(\bk) & = & 
\frac{P^s_{\alpha,\1H}(\bk|R)}{\left[W(k|R)\right]^2} 
=  \frac{1}{\rhob_{\alpha}^2}\int dM n(M)
[W_{\alpha}]^2 \left| U^s(\bk|M)\right|^2 \ ,\label{eq:RedPow1H} \\
P^s_{\alpha,\2H}(\bk) & = & 
\frac{P^s_{\alpha,\2H}(\bk|R)}{\left[W(k|R)\right]^2}  = 
\frac{1}{\rhob_{\alpha}^2}\int \prod_{i=1}^{2} \left\{\frac{}{}dM_i n(M_i)
[W_{\alpha}]_i U^s(\bk|M_i)\right\} 
\frac{P^s_{\hc}(\bk|M_1,M_2,R)}{\left[W(k|R)\right]^2}\
\label{eq:RedPow2H} . \ea
At linear order the redshift space power spectrum of halo seeds is:
\ba
P^s_{\hc}(\bk|M_1,M_2,R) & = &
Z_1(\bk|M_1,R)Z_1(\bk|M_1,R)P_{11}(k) \nonumber \\
& = & [W(kR)]^2 b_1(M_1)b_1(M_2)P_{11}(k) \left\{1+\mu^2 
\left[\beta_1+\beta_2\right]+\beta_1\beta_2 \mu^4 \right\}  
 \ ; \hspace{0.5cm} \beta_i\equiv \frac{f(\Omega)}{b_1(M_i)}\ .
\label{eq:RedPowHaloCent} \ea
where $b_0=0$ can be seen from the fact that the halo and density
fluctuation fields are by definition mean zero fields and recalling
that at linear order
$\left<\delta^h_1(r|M)\right>=b_0(M)+b_1(M)\left<\delta_1(r|M)\right>$.
The redshift space power spectrum monopole is thus
\ba \hat{P}^{s}_{\alpha,\1H}(k) & = & 
\frac{1}{\rhob^2_{\alpha}} \int  dM n(M) [W_{\alpha}]^2
\left|U^{\alpha}(k|M)\right|^2 
{\mathcal R}^{(0)}_{1,2}[k\sigma(M)]\label{eq:Pow1HRedIso} \\
P^{s}_{\alpha,\2H}(k) & = & \frac{1}{\rhob^2_{\alpha}} 
\int\prod_{i=1}^{2} \left\{dM_i
n(M_i)b_1(M_i)[W_{\alpha}]_i U(k|M_i)\right\}
{\mathcal R}^{(0)}_{2,2}[k\sigma_2(M)] \label{eq:Pow2HLegendre}\ea
where we have defined the redshift space multipole factors 
\ba
{\mathcal R}^{(l)}_{1,n}[a] & = & \frac{2l+1}{2}
\int_{-1}^{1} d\mu {\mathcal P}_{l}(\mu) 
\exp\left[-a^2\mu^2\right]\ ;\\
{\mathcal R}^{(l)}_{2,n}[b] & = & \frac{2l+1}{2}
\int_{-1}^{1} d\mu {\mathcal P}_{l}(\mu) 
\left[1+A\mu^2+B\mu^4 \right] \exp\left[-b^2\mu^2\right] \ .
\ea
Here $a^2\equiv nk^2\sigma^2(M)/2$, 
$ ;\ b^2\equiv k^2\left[ \sigma^2(M_1)+\dots+\sigma^2(M_n)\right]/2$
we have set $A=\beta_1+\beta_2$ and $B=\beta_1\beta_2$, and we have
assumed our Gaussian model for the 1-pt velocity distribution function
from Eq.~(\ref{eq:1ptveldist}).  Thus, the monopole $[l=0;\,{\mathcal
P}_0(\mu)=1]$ moments are
\ba
\mathcal{R}^{(0)}_{1,n}(a) & = &
\frac{\sqrt{\pi}}{2}\frac{\erf[a]}{a}\ \ ;\label{eq:PowRed1HMonopole}
\\
\mathcal{R}_{2,n}^{(0)}(b) & = & \frac{\mathcal{R}^{(0)}_{1,n}(b)}{4b^4}
\left[4b^4+2b^2A+3B\right]-
\frac{\exp(-b^2)}{4b^4}\left[2b^2(A+B)+3B\right]
\ \label{eq:Pow2HMonopole} 
\ea
Our expression differs from that of \cite{White2001,Seljak2001}, but
is consistent with the formulation of \cite{Kangetal2002}.  For
further discussion and comments on this subject, and for an evaluation
of the power spectrum to higher order in the Halo-PT series, see
\cite{Smithetal2008}.

Note that when $a\ll 1$ and $b\ll 1$, then 
\ba
\mathcal{R}^{(0)}_{1,n}(a)
   & = & \exp(-a^2)\sum_{j=0}^\infty \frac{2^j}{(2j+1)!!}\,a^{2j}
     =   \sum_{j=0}^\infty \frac{(-1)^j}{j!(2j+1)}\,a^{2j}
\\
\mathcal{R}_{2,n}^{(0)}(b) & = & \frac{\exp(-b^2)}{4b^4}
 \left(\left[4b^4+2b^2A+3B\right]\sum_{j=0}^\infty \frac{2^j}{2j+1)!!}\,b^{2j}
  - 2b^2(A+B) - 3B\right) \nonumber\\
 & = & \frac{\exp(-b^2)}{4b^4} 
 \left(\left[4b^4+2b^2A+3B\right]
 \left[1 + \frac{2b^2}{3} + \frac{4b^4}{15} + \frac{8b^6}{105} + \ldots\right]
  - 2b^2(A+B)- 3B\right) \nonumber\\
 & = & \exp(-b^2)
 \left(1 + \frac{A}{3} + \frac{B}{5} 
       + \frac{2b^2}{3} \left[1 + \frac{A}{5} + \frac{3B}{35}\right] 
       + \frac{4b^4}{15}\left[1 + \frac{A}{7} + \frac{B}{21}\right]
       + \ldots\right)   \nonumber\\
 & \approx & \left(1 + \frac{\beta_1+\beta_2}{3} + \frac{\beta_1\beta_2}{5} \right)
\ea
reducing to the Kaiser formula on large scales \cite{Kaiser1987}.


\section{Impact of shot noise on the reduced bispectrum}\label{app:shotnoise}

It is of interest to consider how standard shot noise and also
the halo model effective shot-noise terms impact the reduced
bispectrum. On large scales $Q^{\rm HM}$ can be written:
\be 
Q^{\rm HM}=\frac{B^{\rm PT}+\frac{1}{\nbar_{\rm 2H,B}}
\left[P_1+P_2+P_3\right]+\frac{1}{\nbar^2_{\rm 1H,B}}}
{\left(P_1+\frac{1}{\nbar_{\rm 1H,P}}\right)\left(P_2+
\frac{1}{\nbar_{\rm 1H,P}}\right)+2\cyc}\ ,
\label{eq:shotnoise}\ee
where we have added a sub-script $P$ or $B$ to distinguish between shot
noise terms from the halo model power spectrum and bispectrum,
respectively.  

In the low sampling limit $\nbarh P\ll 1$, we have the result that 
\be Q^{\rm
HM}=\frac{1}{3}\left[\frac{\left<[W_{\alpha}]^3\right>\left<[W_{\alpha}]\right>}
{\left<[W_{\alpha}]^2\right>^2}\right]\ee
for the case of standard shot noise, the term in square brackets is
unity and we have $Q^{\rm d}=1/3$, where super-script d means
discrete.

In the high sampling limit, $\nbarh P\gg 1$, the denominator in
Eq.~\ref{eq:shotnoise} becomes,
\be \approx \frac{1}{Q_{\rm fac}}\left[1-\frac{2}{\nbar_{\1H,P}}
\frac{P_1+P_2+P_3}{Q_{\rm fac}}-
\frac{3}{\nbar_{\1H,P}^2}\frac{1}{Q_{\rm fac}}\right]\ , \ee
where $Q_{\rm fac}\equiv P_1P_2+2\cyc$ and where we have treated the
last two terms in the square brackets as small quantities, relative to
$Q_{\rm fac}$. On replacing this in Eq.~(\ref{eq:shotnoise}), we find
\be Q^{\rm HM}-Q^{\rm PT}  = \frac{P_1+P_2+P_3}{Q_{\rm fac}}
\left[\frac{1}{\nbar_{\2H,B}}-\frac{2}{\nbar_{\1H,P}}(1+Q^{\rm PT})\right] 
+\frac{1}{Q_{\rm fac}}\left[\frac{1}{\nbar^2_{\1H,B}}-
\frac{3}{\nbar^2_{\1H,P}}(1+Q^{\rm PT})\right] \ .\label{eq:Qshot}\ee
%
We can also use the above expression to derive the effect of standard
shot noise on the reduced bispectrum, by simply considering all of the
number density terms to be identical, and this gives
%
\ba Q^{\rm d}-Q^{\rm PT} =  - \frac{(P_1+P_2+P_3)}{\nbar Q_{\rm fac}}
(1+2Q^{\rm PT})- \frac{1}{\nbar^2 Q_{\rm fac}}(2+3Q^{\rm PT})
 \ .\label{eq:Qstandardshot}\ea
Considering standard shot noise first, in
Eq.~(\ref{eq:Qstandardshot}), we see that the effect of discretization
of matter is always to reduce the value of $Q^d$ relative to $Q^{\rm
PT}$ (the continuum limit case).  Considering now the Halo Model, in
Eq.~(\ref{eq:Qshot}), we see that the difference between this and
$Q^{\rm PT}$ depends on the sign of the quantities in square
brackets. For dark matter, the first term can be seen to be negative,
since $\nbar_{\rm 2H,B}=\nbar_{\rm 1H,P}$. However, the sign of the
second term is not as obvious to deduce. If it is negative, then the
effect is as for standard shot-noise; on the other hand, if the
reverse is true, then $Q^{\rm d}>Q^{\rm PT}$. 

Lastly, the configuration dependence of $Q^{\rm d}-Q^{\rm PT}$ in the
standard shot noise case can be understood from the following: if we
assume that all $k$-vectors are larger than the turn-over scale in the
power spectrum, then the quantity $Q_{\rm fac}(\theta_{12}=1)>Q_{\rm
fac}(\theta_{12}=0)$. This implies that the difference is largest when
$k_1$ and $k_2$ are parallel. 

\vspace{1cm}


\section{Bispectrum Code Comparison}\label{app:codecomp}


\begin{figure*}
\centerline{
\includegraphics[width=6.cm]{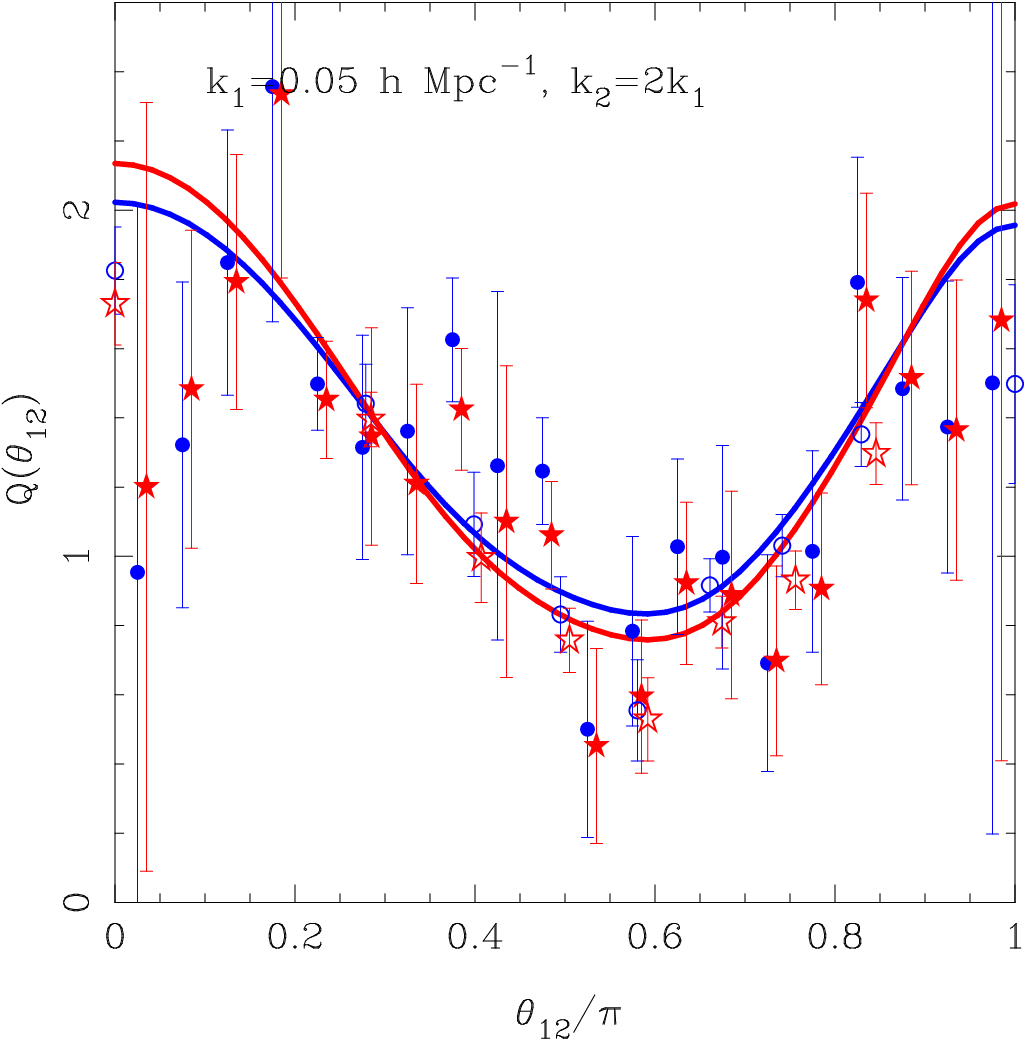}
\includegraphics[width=6.cm]{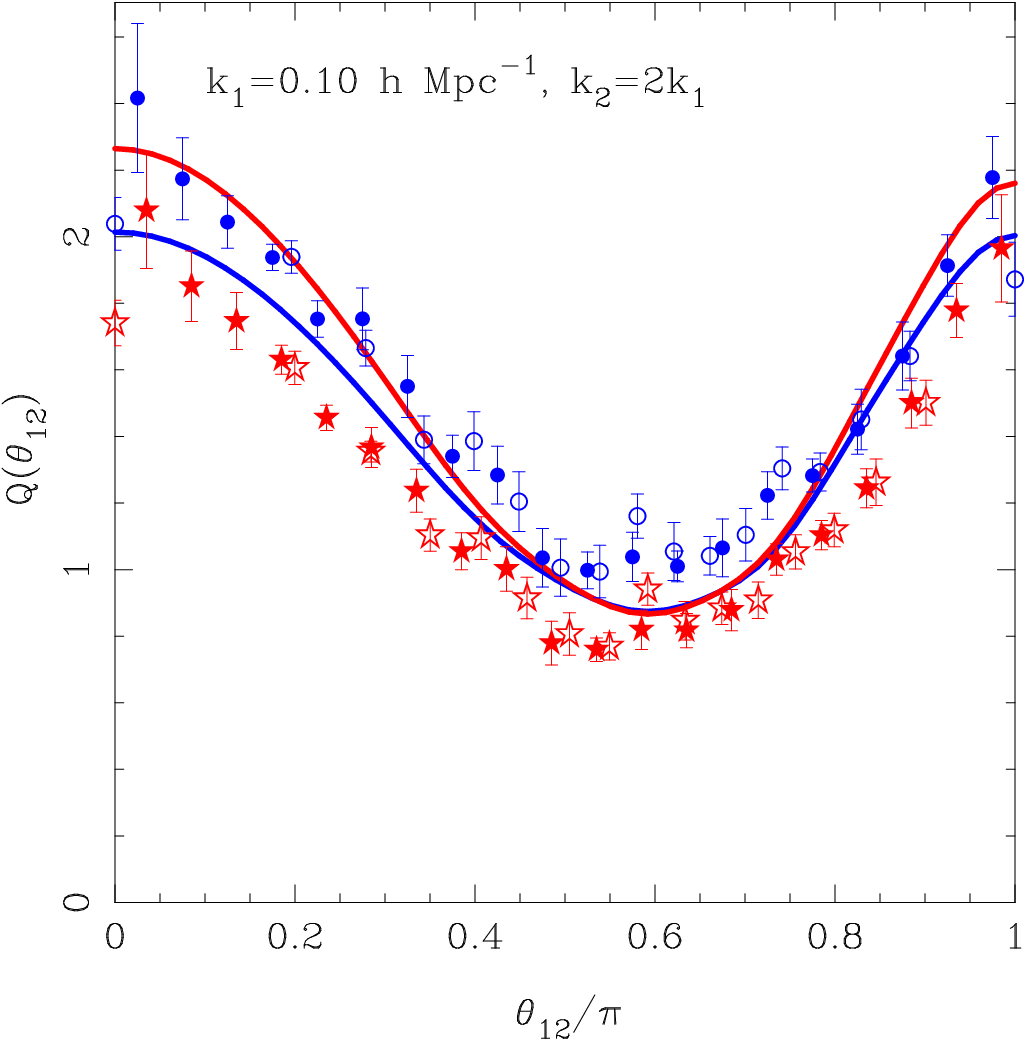}}
\centerline{
\includegraphics[width=6.cm]{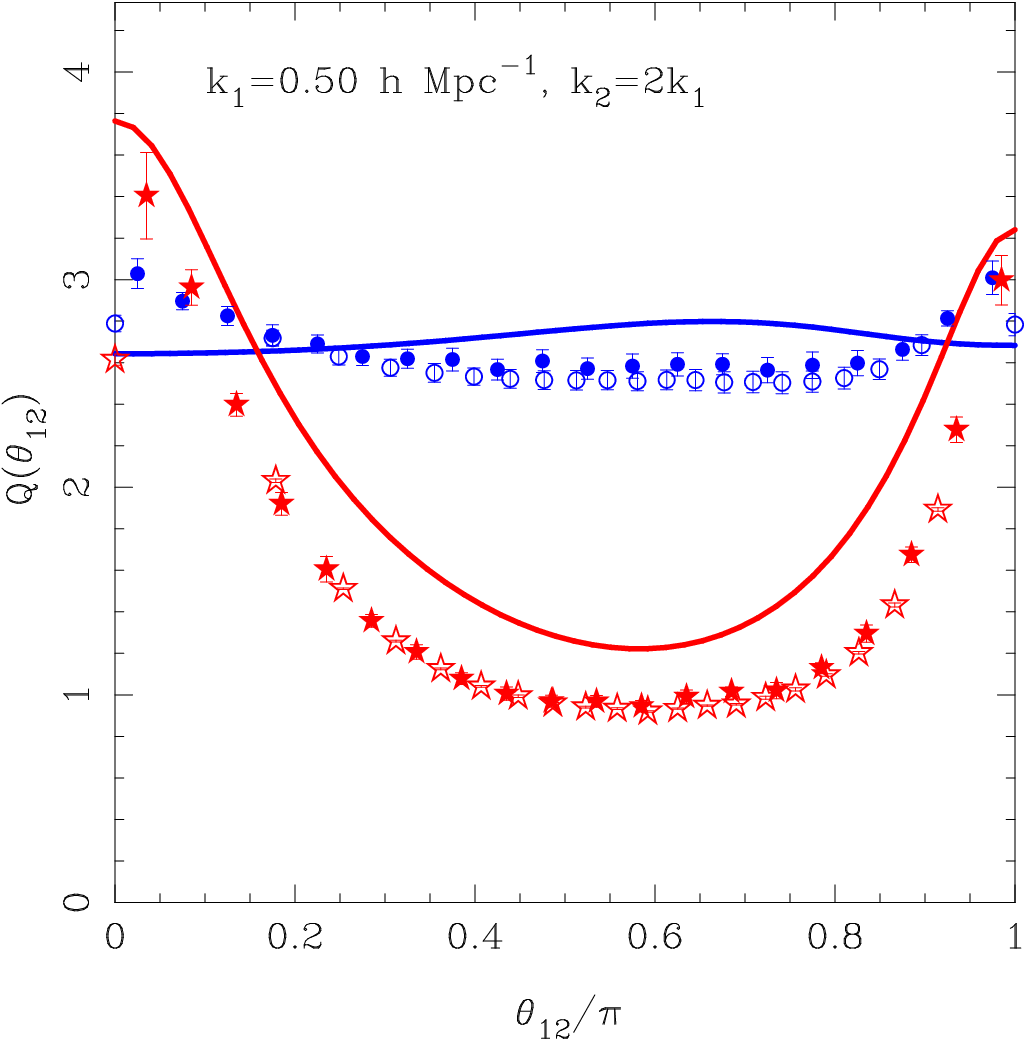}
\includegraphics[width=6.cm]{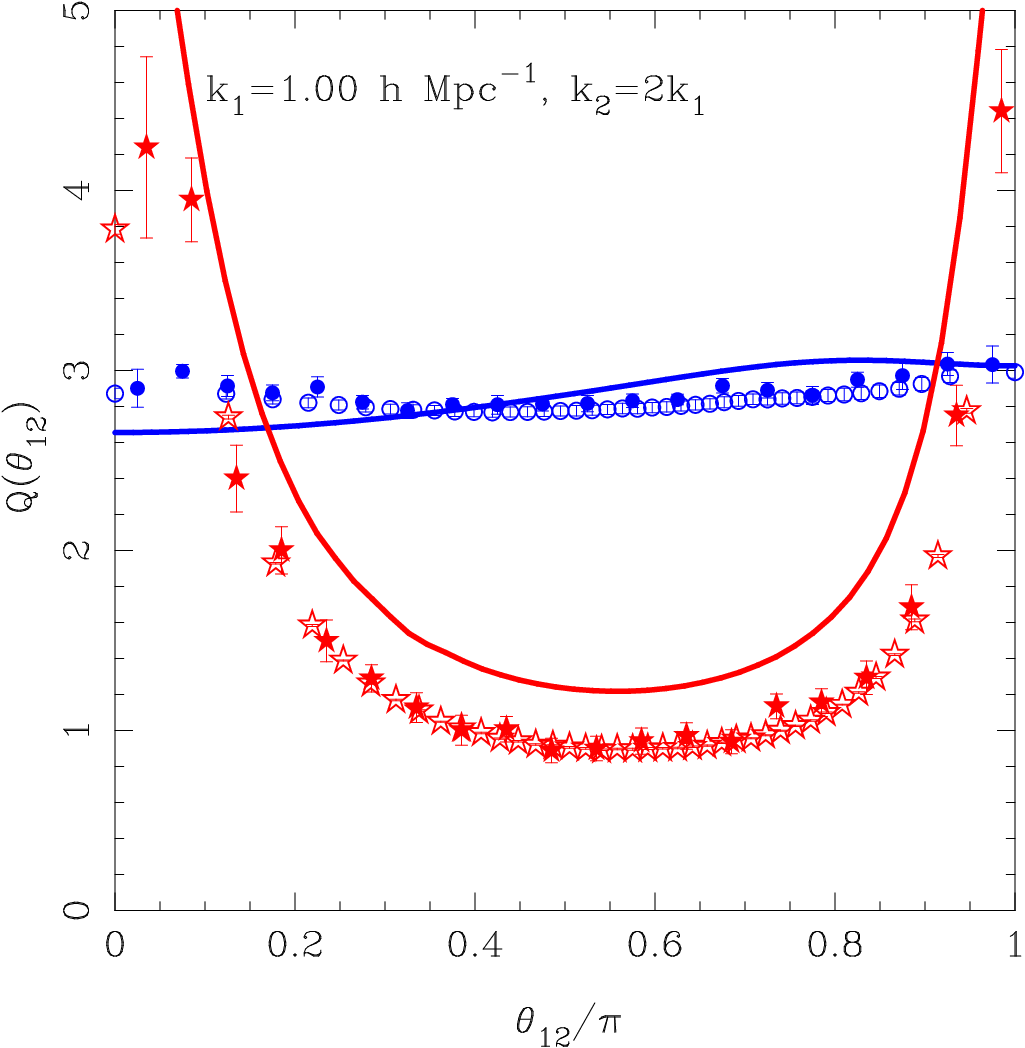}}
\caption{\small{Comparison of the reduced bispectrum estimates
obtained from the method presented in Sec.~\ref{ssec:estimation} with
those obtained from the ``full sampling code'' of
\cite{Scoccimarro2000}. As in Fig.~\ref{fig:QthetaRed3}, our results
are represented by solid symbols, and those from the alternate method
are denoted by open symbols. Again, errors are on the mean and real
and redshift space estimates from the same code have been slightly
offset in the $x$-axis to enhance clarity.}
\label{fig:QthetaRed4}}
\end{figure*}


In this appendix we compare results from our recipe for estimating the
reduced bispectrum, presented in Sec.~\ref{ssec:estimation}, with
those obtained from an independent prescription used by one of us over
the years, e.g. \cite{Scoccimarro2000}, which uses full sampling of
all $k$-space triangles on the Fourier grid. The two methods are very
similar, but some subtle differences exist, these can be summarized:
for the ``full sampling code'':

\begin{enumerate}
\item $B$ is estimated in linear bins of thickness a $1\times k_f$ for
$k={0.05,0.1}[h\Mpc^{-1}]$ and $4\times k_f$ for the case
$k={0.5,1.0}[h\Mpc^{-1}]$ ($k_f=2\pi/L$); whereas for our code
estimates are made in $\Delta\log_{10} k$ bins of thickness $0.05$;
\item the configuration dependence of $B$ is estimated as a linear
function of $k_3$, whereas for our code it is estimated as a linear
function of $\theta_{12}$; 
\item all of the available independent $k$-modes are used, whereas we
sub- or over-sample modes from the available set depending on the
number of available modes;
\item $Q$ is constructed from estimates of $B$ and $P$ in a
post-processing fashion, whereas we estimate $Q$ on-the-fly for each
triangle that is used in the estimate.
\end{enumerate}

Fig.~\ref{fig:QthetaRed4} shows the results of this comparison. The
solid symbols denote our results, and the corresponding open symbols
denote the results from the ``full sampling code''. Overall we find
very good agreement between both methods. On the largest scales that
we have considered, $k_1=0.05\,h\Mpc^{-1}$, it appears that our method
is a factor of 2-3 times more noisy than that of R.~Scoccimarro,
however we have a factor of 2 more bins in $\theta_{12}$, which
accounts for some of this discrepancy. On smaller scales
$k_1\ge0.1\,h\Mpc^{-1}$ the estimates are of comparable quality, with
ours being slightly more noisy. The discrepancy on large scales owes
to the fact that we have sub-sampled triangles from the possible set,
this can be mitigated by oversampling from the number of available
modes. On smaller scales the benefits of our approach are that we may
obtain a high accuracy estimate without requiring all of the triangles
and this also has the practical advantage of keeping the computational
time tolerably low.


\end{widetext}



\def\MNRAS{{Mon. Not. R. Astron. Soc.~}} 
\def\PRD{{Phys. Rev. D.~}}
\def\PRL{{Phys. Rev. Lett.~}} 
\def\ApJ{{Astrophys. J.~}}
\def\ApJS{{Astrophys. J.~Supp.}}
\def\AA{{Astron. Astrophys.~}} 
\def\Nat{{Nature (London)~}}
\def\AstroPart{{Astro-particle Phys.~}}
\def\AJ{{Astron. J.~}}
\def\PASJ{{Publ. Astron. Soc. Japan}}


\end{document}